\documentclass{article}
\usepackage[a4paper, portrait, margin=1.1811in]{geometry}
\usepackage[english]{babel}
\usepackage[utf8]{inputenc}
\usepackage[T1]{fontenc}
\usepackage{helvet}
\usepackage{etoolbox}
\usepackage{graphicx}
\usepackage{titlesec}
\usepackage[font=footnotesize]{caption}
\usepackage{subcaption}
\usepackage{booktabs}
\usepackage{xcolor} 
\graphicspath{ {./images/} }
\usepackage{scrextend}
\usepackage{fancyhdr}
\usepackage{graphicx}
\newcounter{lemma}

\newcounter{theorem}

\usepackage[export]{adjustbox}
\usepackage{amsmath}
\usepackage{breqn}
\DeclareMathOperator{\rect}{rect}
\DeclareMathOperator{\sinc}{sinc}
\usepackage[backend=biber,sorting=none, minbibnames=3]{biblatex}
\usepackage{array}
\usepackage{bm}
\usepackage{placeins}
\usepackage{csquotes}
\usepackage{float}
\usepackage[colorlinks, citecolor=cyan]{hyperref}
\usepackage{bookmark}
\usepackage{hhline}

\fancypagestyle{plain}{
	\fancyhf{}

}

\makeatletter
\patchcmd{\@maketitle}{\LARGE \@title}{\fontsize{16}{19.2}\selectfont\@title}{}{}
\makeatother

\usepackage{authblk}

\newsavebox\affbox
\author[1]{\textbf{Hunter Meyer}}
\author[1]{\textbf{Joyoni Dey}}
\author[1]{\textbf{Sydney Carr}}
\author[2]{\textbf{Kyungmin Ham}}
\author[3]{\textbf{Leslie G. Butler}}
\author[4]{\textbf{Kerry M. Dooley}}
\author[1,5]{\textbf{Ivan Hidrovo}}
\author[6]{\textbf{Markus Bleuel}}
\author[7]{\textbf{Tamas Varga}}
\author[8,9]{\textbf{Joachim Schulz}}
\author[8]{\textbf{Thomas Beckenbach}}
\author[8]{\textbf{Konradin Kaiser}}

\affil[1]{ Department of Physics and Astronomy, Louisiana State University,
	Baton Rouge, LA, 70803
}
\affil[2]{ Center for Advanced Microstructures and Devices, Louisiana State University,
    Baton Rouge, LA, 70806
}

\affil[3]{ Department of Chemistry, Louisiana State University, 
	Baton Rouge, LA, 70803
 }
\affil[4]{ Cain Department of Chemical Engineering,  Louisiana State University,
Baton Rouge, LA, 70803
}
\affil[5]{Department of Radiation Therapy, Solón Espinosa Ayala Oncological Hospital, Quito, Ecuador.}

\affil[6]{ Adelphi Technology, Inc., Redwood City, CA, 94063
}

\affil[7]{ The Environmental Molecular Sciences Laboratory, Pacific Northwest National Laboratory, Richland, WA}

\affil[8]{ Microworks GmbH, Schnetzlerstr. 9, 76137 Karlsruhe,
Germany
}
\affil[9]{Institute of Microstructure Technology, Karlsruhe Institute of Technology,
Hermann-von-Helmholtz-Platz 1, D-76344
Eggenstein-Leopoldshafen, Germany. }

\titlespacing\section{0pt}{12pt plus 4pt minus 2pt}{0pt plus 2pt minus 2pt}
\titlespacing\subsection{12pt}{12pt plus 4pt minus 2pt}{0pt plus 2pt minus 2pt}
\titlespacing\subsubsection{12pt}{12pt plus 4pt minus 2pt}{0pt plus 2pt minus 2pt}

\titleformat{\section}{\normalfont\fontsize{12}{15}\bfseries}{\thesection.}{1em}{}
\titleformat{\subsection}{\normalfont\fontsize{12}{15}\bfseries}{\thesubsection.}{1em}{}
\titleformat{\subsubsection}{\normalfont\fontsize{12}{15}\bfseries}{\thesubsubsection.}{1em}{}

\titleformat{\author}{\normalfont\fontsize{12}{15}\bfseries}{\thesection}{1em}{}

\title{\textbf{\huge X-ray Interferometry Using a Modulated Phase Grating: Theory and Experiments}\\
	}
\date{}    

\begin{document}

\newpage
\setcounter{page}{1}
\renewcommand{\thepage}{\arabic{page}}

\captionsetup[figure]{labelfont={bf},labelformat={default},labelsep=period,name={Figure}}
\captionsetup[table]{name=Table, labelfont=bf, labelsep=period}

\setlength{\parskip}{0.5em}

\maketitle

\noindent\rule{15cm}{0.5pt}
\begin{abstract}

X-ray grating interferometry allows for the simultaneous acquisition of attenuation, differential-phase contrast, and dark-field images, resulting from X-ray attenuation, refraction, and small-angle scattering, respectively.  The modulated phase grating (MPG) interferometer is a recently developed grating interferometry system capable of generating a directly resolvable interference pattern using a relatively large period grating envelope function that is sampled at a pitch that allows for X-ray spatial coherence using a microfocus X-ray source or by use of a source G0 grating that follows the Lau condition.  We present the theory of the MPG interferometry system for a 2-dimensional staggered grating, derived using Fourier optics, and we compare the theoretical predictions with experiments we have performed with a microfocus X-ray system at Pennington Biomedical Research Center, LSU.  The theoretical and experimental fringe visibility is evaluated as a function of grating-to-detector distance. Quantitative experiments are performed with porous carbon and alumina samples, and qualitative analysis of attenuation and dark-field images of a dried anchovy are shown.

\textbf{\textit{Keywords}}: \textit{X-ray interferometry ; modulated phase grating ; diffraction grating; dark-field; porosity}
\end{abstract}
\noindent\rule{15cm}{0.4pt}

\section{Introduction}

X-ray grating interferometry allows for the simultaneous acquisition of attenuation, differential-phase contrast (DPC), and dark-field images.  In contrast to traditional X-ray radiography systems, one or more diffraction gratings are placed in the path of the X-ray beam so that a periodic interference pattern is produced, commonly referred to as interference fringes, which is approximately sinusoidal. With no object in the path of the X-ray beam, the interference pattern is typically referred to as the reference or \textit{blank} image.  When an object is placed in the path of the X-ray beam, it's physical properties are imaged by measuring the perturbation to the reference fringe pattern. The pattern is perturbed in three ways, resulting in images with three distinct contrast mechanisms \cite{bib:Pfeiffer2009}.  Attenuation causes a reduction in the average value of the fringe pattern, producing the attenuation image.  Refraction results in a phase shift of the pattern, producing the DPC image.  Small angle scattering reduces a parameter known as fringe visibility, which is simply the height of the fringes relative to the average value, producing dark-field images.  Interferometry has potential for a variety of applications in science and medicine, including lung imaging \cite{bib:Gassert, bib:Bech, bib:Yaroshenko, bib:Velroyen}, breast imaging \cite{bib:Wang2014, bib:Tapfer, bib:Scherer, bib:Koehler}, arthritis imaging \cite{bib:Stutman, bib:Tanaka}, osteoporosis imaging \cite{bib:Gassert2023}, pore size analysis \cite{bib:Revol}, additive manufacturing quality assurance \cite{bib:Zhao, bib:Brooks}, etc.

There are presently several grating interferometers in the literature, including the Talbot-Lau Interferometer (TLI) \cite{bib:Momose2003,bib:Momose2005,bib:Pfeiffer2006}, Dual Phase Grating Interferometer (DPGI) \cite{bib:Miao, bib:Yan}, and Modulated Phase Grating Interferometer (MPGI) \cite{bib:MPGPatent1,bib:MPGPatent2,bib:ParkXuDey,bib:HidrovoMeyerRSI,bib:PandeshwarStampanoni}.  The TLI has a phase grating, G1, that produces interference fringes that are not directly resolvable by typical detectors, meaning an analyzer grating, G2, is required to create visible Moiré patterns, resolvable at the detector.  The analyzer grating's pitch is determined by the geometric magnification of the G1 grating, meaning the geometry of the system is fixed.  While this interferometer is highly sensitive due to the low-period fringes produced, the requirement of an analyzer grating, which is an absorption grating, doubles the dose per image for similar fluence at the detector.  The DPGI achieves directly resolvable patterns using two phase gratings separated by a few millimeters, without the need for an analyzer grating.  The Moiré pattern produced has a beat pattern directly resolvable by the detector, with the high-frequency components being washed out by blur from the detector and X-ray source.  The DPGI configurations are typically far-field geometries and are often called a Far-Field interferometry systems \cite{bib:Miao}. The DPGI's two phase gratings must be co-aligned for proper fringe formation, which often requires time-consuming alignment procedures.  This problem exists for the TLI as well.

\begin{figure}
    \centering
    \includegraphics[keepaspectratio = true, width = 0.7\textwidth]{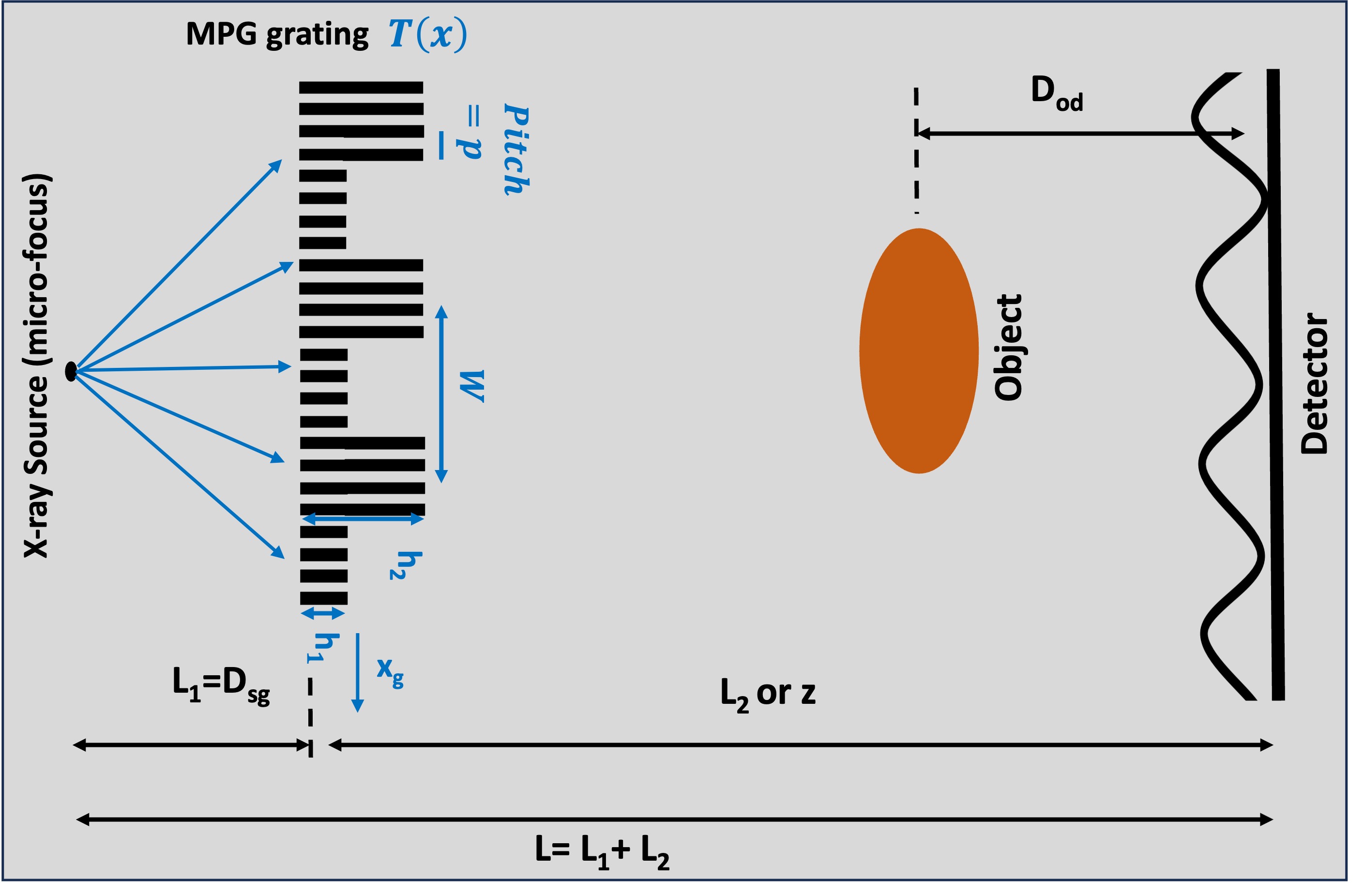}
    \caption{Schematic of the modulated phase grating interferometer with a micro-focus X-ray source.  Here the 1-dimensional RectMPG parameters are shown, where $W$ is the period of the grating's envelope function, $p$ is the grating pitch, $h_1$ and $h_2$ are the phase heights, and the geometry is defined by the source-to-grating distance, $L_1$, the grating-to-detector distance, $z$, and the object-to-detector distance, $D_{od}$.  $x_g$ denotes the phase stepping direction.}
    \label{fig:MPG_schematic}
\end{figure}

The modulated phase grating interferometer (MPGI), originated by our group \cite{bib:MPGPatent1,bib:MPGPatent2,bib:ParkXuDey,bib:HidrovoMeyerRSI}, consists of a single grating where the heights of the grating bars follows an envelope function, with a relatively large period, $W$.  The envelope function is sampled at a high-frequency pitch, $p$, to meet the coherence requirements similar to that of the TLI and DPGI systems.  In this context, the envelope period, $W$, and the sampling pitch, $p$, should not be confused, since this terminology is used interchangeably for a binary diffraction grating such as those in the TLI or DPGI systems.  The fringes produced have many high-frequency harmonics that are washed out by the detector and low-frequency harmonics that result from the envelope function, analogous to the beat pattern produced by the DPGI.  Images can be calculated using either a single-shot \cite{bib:Bevins, bib:Wen2010} or a phase stepping procedure \cite{bib:Marathe}.  While the single-shot methodology is simpler, the fringe resolution is not as high, so for this study, phase stepping was used.

A key benefit of the MPGI is the simplicity of using a single grating, with system geometry shown in Figure \ref{fig:MPG_schematic}.  Since no analyzer grating is present, MPG placement is not limited by Talbot distance constraints as in the TLI.  The MPG can be moved continuously, allowing for a range of system autocorrelation lengths.  In the TLI, the autocorrelation length can only be changed by moving the object, so an extra degree of freedom is introduced for the MPGI.  For the DPGI, the autocorrelation length can be changed by changing the inter-grating distance, but grating alignment must be maintained.  Another more subtle advantage is that the MPGI system is always first harmonic dominant.  In contrast, the TLI and DPGI systems are often second harmonic dominant; the first harmonic disappears if they are illuminated by a monochromatic source.  The first harmonic reappears when illuminated by a polychromatic source \cite{bib:Yan}, causing the interference fringes to not be as sinusoidal, leading to ambiguities in the systems' phase sensitivity, autocorrelation length, and visibility.

In this study, we present the theory of fringe formation for the MPGI for a polychromatic microfocus X-ray tube source.  We then compare the results with experiments performed in the Keck Imaging Laboratory at Pennington Biomedical Research Center (PBRC). In a previous study, we presented the theory of the 1-dimensional MPG with comparisons to simulations performed using the Sommerfeld Rayleigh Diffraction Integral (SRDI) simulator, with several orders of magnitude of decreased simulation time and comparable results \cite{bib:HidrovoMeyerRSI}.  In this paper, we will extend the 1-dimensional MPG theory to account for the staggering of the grating bars that are implemented during fabrication for grating stability.

The presented theory models diffraction from the phase and attenuation of the grating bars using Fourier optics, as well as the effects of using a polychromatic source, finite focal spot size, and detector point spread function (PSF).  The theory allows for the rapid calculation of the fringe profile and fringe visibility as a function of system geometry and MPG parameters.  The theory can also be used to aid in developing future MPG interferometry systems by allowing for rapid visibility calculations for a wide range of system geometries, energies, and grating designs. The experiments performed at PBRC include a measurement of the fringe visibility as a function of grating-to-detector distance and images taken of a variety of samples.  Experiments were performed using rectangular envelope MPGs, known as RectMPGs, with a schematic shown in Figure \ref{fig:MPG_schematic}. Additionally, we measured the mean attenuation and dark-field signal for several porous carbon and alumina samples and acquired images of a dried anchovy.

\section{Methods}
\subsection{Theory}
\label{subsec:theory}

\begin{figure}
    \centering
    \includegraphics[keepaspectratio = true, width = 0.7\textwidth]{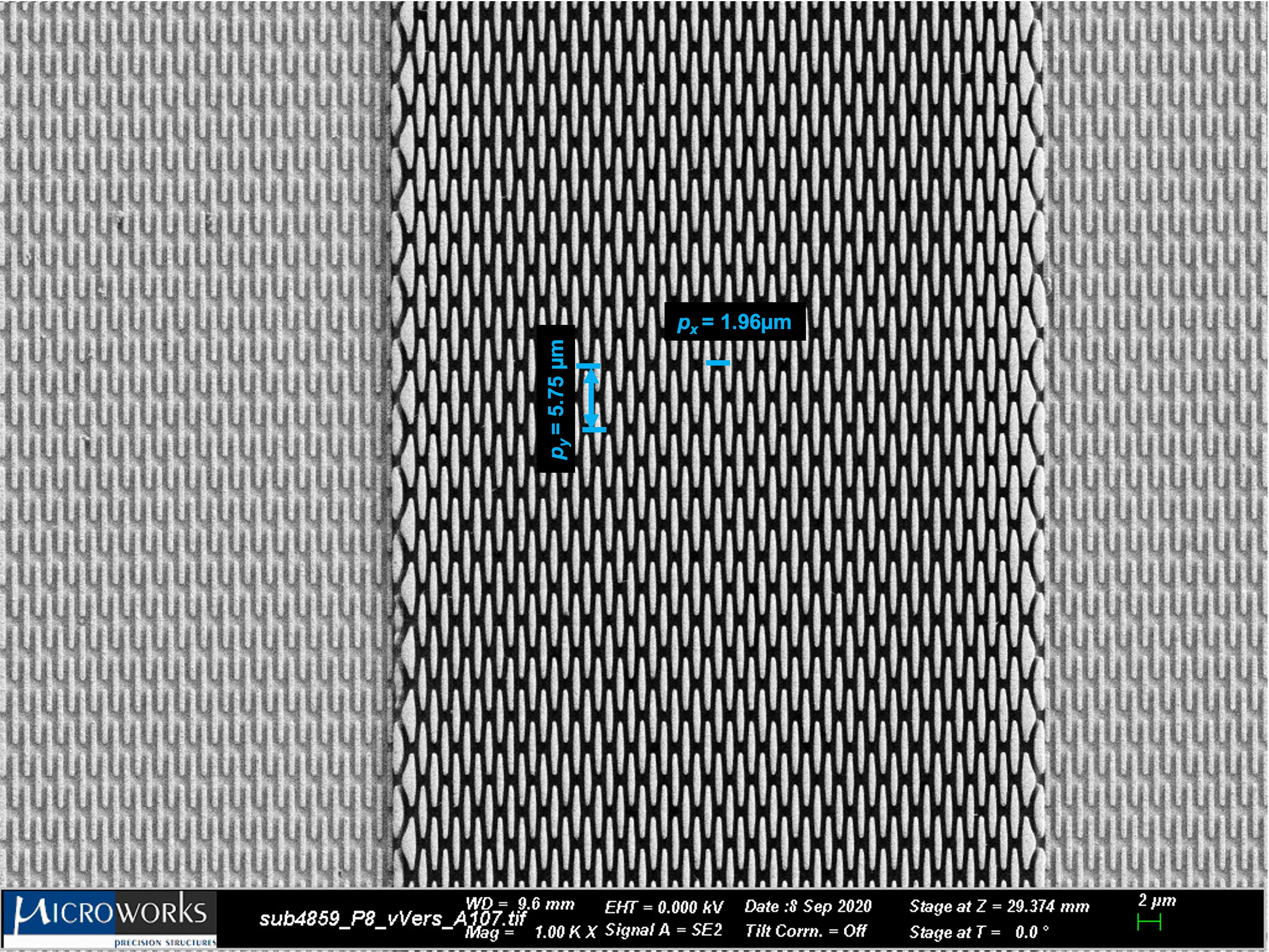}
    \caption{An SEM image of the staggered 2D modulated phase grating, with $p_x$ and $p_y$ labeled.  $\alpha_x$ and $\alpha_y$ are the duty cycles of the grating in the x- and y-dimension, respectively.}
    \label{fig:staggered_MPG}
\end{figure}

The theory of the 1-dimensional MPG was originally derived in \cite{bib:HidrovoMeyerRSI}, which will hold for a 2-dimensional MPG if the grating is constant in one dimension.  However, for this study, multiple MPGs were produced by Microworks GmbH, Germany, where the grating bars were staggered for improved structural stability, following a Bridge design, \cite{bib:Trimborn}, as shown in Figure \ref{fig:staggered_MPG}.  Here we present a revised version of the MPG theory that accounts for the staggered grating bars.  This study will compare the theoretically calculated visibility with experimental measurements of fringe visibility as a function of grating-to-detector distance.  

\subsubsection{A Modified Transmission Function in 1 Dimension}

First, we will briefly review the theory of the 1-dimensional MPG \cite{bib:HidrovoMeyerRSI}, then detail the modifications necessary for 2 dimensions.  For the 1-dimensional MPG, the grating transmission function is given as

\begin{gather}
\label{eq:old_1D_transmission}
T(x) = \left[ g(x)\sum_{n = -\infty}^{\infty} \delta(x - np) + \sum_{n = -\infty}^{\infty} \delta(x - np - p/2) \right] \star \rect\left(\frac{x}{\alpha p}\right)
\\
g(x) = \sum_{m = \infty}^{\infty} g_m \exp\left(\frac{2\pi j mx}{W}\right)
\end{gather}

The envelope function is represented by $g(x)$ and is periodic.  It can be represented as a Fourier Series of period $W$, with Fourier coefficients, $g_m$.  In Equation \ref{eq:old_1D_transmission}, the first series of $\delta$ functions samples the envelope function at a sampling period of $p$, and the convolution with the $\rect$ function places a grating bar at each sample, since $f(x) \star \delta(x - x_0) = f(x - x_0)$, where $\star$ represents the convolution.  $\alpha$ represents the duty cycle of the grating and $j$ is the imaginary number.  The second series of $\delta$ functions is responsible for the gaps between the grating bars, where the grating transmission equals 1, not 0.

The envelope function's Fourier coefficients, $g_m$, are energy-dependent and account for the transmission function's amplitude reduction and phase change.  The physical heights of the grating bars are designed to follow the desired phase envelope at the design energy, $E_D$.  The grating attenuation is determined by the physical heights of the grating bars resulting from that design.  The attenuation and phase profile of the grating are energy-dependent, therefore we represent the grating coefficients as $g_m(\lambda)$.

The polychromatic intensity is simply the incoherent superposition of each monochromatic intensity weighted by the energy spectrum.  This means the energy spectrum affects the detector intensity in two ways: by diffraction and by the grating coefficients.  The field amplitude can be derived using the angular spectrum method under the Fresnel approximation (Goodman \cite{bib:Goodman}).  The detector intensity can be calculated as simply the square of the field amplitude, $I(x,z) = |U(x,z)|^2$.

It is difficult to generalize Equation \ref{eq:old_1D_transmission} to 2 dimensions, since defining the regions between the grating bars becomes cumbersome.  Instead, an equivalent transmission function, written in another form, is introduced. Here, the grating bars result from sampling, and a constant transmission of 1 is included to account for the regions between the grating bars.  This constant transmission is then subtracted from $g(x)$ so that the transmission of the grating bars follows the envelope function.  The equivalent transmission function in 1 dimension is shown in Equation \ref{eq:new_1D_transmission}. It is straightforward to show that this function is exactly the same as Equation \ref{eq:old_1D_transmission}.  

The new transmission function is used to derive the field amplitude shown in Equation \ref{eq:new_1D_fieldAmplitude} using the angular spectrum method \cite{bib:Goodman}. Though there are notational differences, the calculated field amplitude is exactly the same as what is found in our previous work \cite{bib:HidrovoMeyerRSI} (using the first transmission function Equation \ref{eq:old_1D_transmission}).

\begin{gather}
\label{eq:new_1D_transmission}
T(x) = 1 + \left[ (g(x) - 1) \sum_{n = -\infty}^{\infty} \delta(x - np) \right] \star rect\left(\frac{x}{\alpha p}\right) \\
\label{eq:new_1D_fieldAmplitude}
U(x,z) = 1
+ \alpha \sum_m \sum_n b_1(m, n, z) \exp\left( j 2 \pi x \left( \frac{m}{W} + \frac{n}{p} \right) \right)
- \alpha \sum_n b_2(n, z) \exp\left(j 2 \pi x \left(\frac{n}{p}\right)\right)
\\
b_1(m,n, z) = g_m \sinc\left( \alpha p \left( \frac{m}{W} + \frac{n}{p} \right) \right) \exp\left(-j \pi \lambda z \left( \frac{m}{W} + \frac{n}{p} \right)^2 \right)
\\
b_2(n, z) = \sinc\left(\alpha n\right) \exp\left(-j \pi \lambda z \left(\frac{n}{p}\right)^2\right)
\end{gather}

\subsubsection{The 2-Dimensional Transmission Function}

Equation \ref{eq:new_1D_transmission} is adapted to two-dimensions, shown in Equation \ref{eq:2D_transmission}.  A noticeable change is the reintroduction of a second sampling function, (not to be confused with the double sampling in Equation \ref{eq:old_1D_transmission}).  Additionally, there is now a pitch in each direction, $p_x$ and $p_y$, as well a duty cycle in each direction, $\alpha_x$ and $\alpha_y$.  Here, the two sampling terms both follow $g(x)$ but are offset by $p_x/2$ and $p_y/2$, to account for the staggering of the grating bars.  Note that the staggering is removed if $\alpha_y = 1$, $\alpha_x = \alpha/2$, $p_x = 2 p$, and the 2D transmission function will be equivalent to the 1D transmission function.

\begin{dmath}
\label{eq:2D_transmission}
T(x,y) = 1 + \Bigg\{ (g(x) - 1) \biggl[\sum_n \sum_l \delta(x - n p_x)\delta(y - l p_y)
+ \sum_n\sum_l \delta \left(x - n p_x - \frac{p_x}{2} \right) \delta \left(y - l p_y - \frac{p_y}{2} \right)
\biggr] \Bigg\} 
\star rect \left( \frac{x}{\alpha_x p_x}\right) rect \left( \frac{y}{\alpha_y p_y}\right)
\end{dmath}

In Appendix \ref{appen:2D_MPG_derivation}, the field amplitude is derived using the angular spectrum method \cite{bib:Goodman}, followed by the Fresnel scaling theorem \cite{bib:Paganin} to scale the intensity from that of a plane-wave to that of a point-source.  The intensity at the detector is calculated as 

\begin{equation}
    I(x,y,z) = 1 + \sum_l \biggl[ c(l,x,z) + c^*(-l,x,z) + d(l,x,z) \biggr] \exp \left( j 2 \pi y \frac{l}{M p_y} \right)
\end{equation}

\noindent where $c(l,x,z)$ is as defined in Appendix \ref{appen:2D_MPG_derivation} and $d(l,x,z) = c(l,x,z) \star c^*(-l,x,z)$. \textcolor{black}{$c(l,x,z)$ and $d(l,x,z)$ contain the x-harmonics.}

Next, as detailed in Appendix \ref{appen:2D_MPG_derivation} (Equations \ref{eq:1D_approx_intensity}-\ref{eq:dl0}), to simplify calculations, the 2D intensity is approximated by the $l = 0$ harmonic, where $l$ represents the harmonic number in the y-dimension. \textit{This means that we keep only the zero-th harmonic (mean) in the y-direction}. This approximation is valid when the detector and source blur are sufficiently high enough to blur the non-zero y-harmonics that result from the staggering of the grating bars necessary for fabrication stability.  Notably, while this approximation removes the dependence of $p_y$, the fringe profile still depends on the duty cycle in the y-direction $\alpha_y$.  The staggering of the grating bars still affects the fringe visibility, even when there is sufficient blur to approximate the 2-dimensional intensity profile into a single dimension.

\begin{equation}
\label{eq:1D_approx_intensity_reproduced}
    I(x,y,z) \approx 1 + c(0, x, z) + c^*(0, x, z) + d(0, x, z)
\end{equation}

\textcolor{black}{Appendix A Equation 39 is the intensity calculated under the $l = 0$ Approximation and is reproduced in Equation \ref{eq:1D_approx_intensity_reproduced}.}  \textit{The $l = 0$ approximation provides a powerful tool to make 2D intensity calculations fast under some often realistic conditions when the detector point spread function (PSF) and source blur combination is larger than the (magnified) grating pitch in the y-dimension, $p_y$.}  This approximation greatly reduces the computation required, since only one dimension needs to be calculated, instead of two.

The theory is evaluated in several ways in the Results, Section \ref{subsec:theory_results}.  The intensity profile calculated for the 2D MPG under the $l = 0$ approximation is shown to be equivalent to a full 2D MPG simulation with sufficient blur in the y-direction and a sufficiently small $p_y$.  \textcolor{black}{It is also shown that the $l = 0$ approximation is further reduced to the true 1D MPG---where it is constant in one dimension and the field amplitude is shown in Equation \ref{eq:new_1D_fieldAmplitude}---under the conditions of $\alpha_x = \alpha/2$, $p_x = 2p$, $\alpha_y = 1$.}

\subsubsection{Post-Processing and Visibility Calculations}

The intensity derived in Appendix \ref{appen:2D_MPG_derivation} Equation \ref{eq:1D_approx_intensity} is the intensity profile when the source is a monochromatic point source.  To properly model the fringe profile and visibility measured in an experiment, we must include a polychromatic source, finite focal spot size, and detector point spread function.  Additional post-processing includes modeling the phase stepping procedure and downsampling to the detector sample rate.

The intensity of a polychromatic point source can be found by integrating over the energy spectrum, $S(E)$.

\begin{equation}
    I_{poly}(x,y,z) = \int_E I(x,y,z; E) S(E) dE
\end{equation}

The effect of finite focal spot size can be found by convolving the point-source intensity profile with the magnified source profile.  It's important to note that the source is magnified by a factor of $M - 1 = \frac{z}{L1}$, as shown in Appendix B of \cite{bib:HidrovoMeyerRSI},

\begin{equation}
    I_{poly, spot}(x,y,z) = I_{poly}(x,y,z) \star \sigma(\frac{x}{M-1})
\end{equation}

Finally, the detector point spread function (PSF) is convolved to get the final intensity profile,

\begin{equation}
    I_{final}(x,y,z) = I_{poly, spot}(x,y,z) \star PSF(x,y)
\end{equation}

The phase stepping procedure is modeled by shifting the intensity profile by the phase step size, $x_s$, multiplied by the magnification factor, $M$.  Following this, downsampling is performed to match the detector pixel size.  The intensity is initially calculated at a sample rate much higher than the pixel size.  For the purposes of this study, the intensity profile is initially calculated at a rate of $0.1 \: \mu m$, and the Dexela 1512 detector used in this study has a pixel size of $75 \: \mu m$.  It's important to note that the detector PSF and pixel size are not the same.  Additionally, in the theoretical simulations, downsampling should occur \textit{after} phase stepping.  The visibility is measured following the methods of Marathe et al. \cite{bib:Marathe}.

\subsection{Experimental Methods}
\label{subsec:experimental_methods}

Several experiments were performed in the Keck Imaging Laboratory at Pennington Biomedical Research Center (PBRC) for the purposes of this study.  Two MPGs were used, both with rectangular phase modulation and the same design parameters, listed in Table \ref{tab:MPG_parameters}.  \textcolor{black}{The gratings, referred to as MPG7 and MPG8, were manufactured by Microworks, GmbH. They provided the measured heights of the grating structures, listed in Table \ref{tab:MPG_height_measurements}, where it is seen that MPG8 has a larger difference in the heights than MPG7.  Due to the novelty of the grating fabrication process, the height of the grating structures ($h_1$ and $h_2$) is not highly reproducible, which can explain the observed differences.}  A Hamamatsu L9181-02 microfocus X-ray tube was used with a Dexela 1512 X-ray Detector.  The X-ray source was consistently run at $45 \: kVp$ and $55 \: \mu A$, under the small focus spot mode ($5 - 8 \: \mu m$).  

\begin{table}[]
    \centering
    \begin{tabular}{|c||c|}
        \hline
        \textbf{MPG Parameter} & \textbf{MPG7/MPG8} \\
        \hline
        $\boldsymbol{W \: (\mu m)}$ & 120 \\
        \hline
        $\boldsymbol{p_{x} \: (\mu m)}$ & 1.96 \\
        \hline
        $\boldsymbol{p_{y} \: (\mu m)}$ & 5.75 \\
        \hline
        $\boldsymbol{\alpha_{x}}$ & 0.25 \\
        \hline
        $\boldsymbol{\alpha_{y}}$ & 0.86 \\
        \hline
        \textbf{Design} $\boldsymbol{h_{1}}$ & $\pi / 8$\\
        \hline
        \textbf{Design} $\boldsymbol{h_{2}}$ & $\pi / 2$ \\
        \hline
        \textbf{Material} & Gold (Au) \\
        \hline
    \end{tabular}
    \caption{Modulated phase grating parameters used in visibility measurements.  \textcolor{black}{$h_1$ and $h_2$ are listed as their corresponding phase-shift in radians}.}
    \label{tab:MPG_parameters}
\end{table}

\begin{table}[]
    \centering
    \begin{tabular}{|c||c|c|c|c|}
        \hline
         & \textbf{Design} $\boldsymbol{\phi}$ & \textbf{Design Height} $\boldsymbol{(\mu m)}$ & \textbf{MPG7 Height} $\boldsymbol{(\mu m)}$ & \textbf{MPG8 Height} $\boldsymbol{(\mu m)}$ \\
         \hline
         $\boldsymbol{h_{2}}$ & $\pi / 2$ & $2.41$ & $2.17 \pm 0.57$ & $2.50 \pm 0.41$ \\
         \hline
         $\boldsymbol{h_{1}}$ & $\pi / 8$ & $0.602$ & $0.69 \pm 0.17$ & $0.62 \pm 0.17$ \\
         \hline
         $\boldsymbol{\Delta h}$ & $3\pi/8$ & $1.81$ & $1.48$ & $1.88$ \\
         \hline
    \end{tabular}
    \caption{\textcolor{black}{Comparison of the average measured height of the grating structures, $h_1$ and $h_2$, with the designed heights, for MPG7 and MPG8.  Uncertainties shown are the standard deviation. The heights were measured by the manufacturer, Microworks, GmbH.}}
    \label{tab:MPG_height_measurements}
\end{table}

\textcolor{black}{The fringe visibility produced by each grating was measured as a function of grating-to-detector distance, $z$, for a fixed source-to-detector distance, $L = 110 \: cm$.  For the purposes of comparing the experimentally acquired visibility with that predicted by the presented theory, the detector PSF was measured by imaging the TO MTF tungsten edge phantom by Leeds Test Objects \cite{bib:Leeds} at a small angle close to the detector.  A forward model was produced of an angled edge and a generalized Gaussian was fit to find the blur induced by the detector, yielding a PSF with generalized Gaussian parameters of $\sigma = 57.87 \: \mu m$ and $k = 0.9$.  The source profile was measured following the methods of Nishiki et al. \cite{bib:Nishiki}, again using the TO MTF phantom, yielding a Gaussian source profile with $\sigma = 3.14 \: \mu m$.}

\textcolor{black}{Next, images of several porous carbon and alumina samples were acquired over a range of autocorrelation lengths of approximately $20 - 80 \: nm$ and attenuation, dark-field, and DPC images were calculated.  Lastly, an anchovy was imaged at approximately the maximum visibility positions for the two gratings and the attenuation and dark-field images were calculated.  Image parameters were measured using the methods of Marathe et al., \cite{bib:Marathe}, but log-scale attenuation and dark-field images were used in our analysis, as explained in Section \ref{subsec:carbon_and_alumina_methods}.}

\subsubsection{Visibility Measurements}
\label{subsec:exper_vis_measurements}
The visibility was measured as a function of grating-to-detector distance, $z$, for a fixed source-to-detector distance, $L = 110 \: cm$.  Two MPGs,  labeled MPG7 and MPG8, each with rectangular phase modulation, were used.  The MPG parameters are listed in Table \ref{tab:MPG_parameters}.  The grating-to-detector distance ranged from $56 - 98 \: cm$ in $2 \: cm$ increments.  The grating was phase stepped 7 times, at $30 \: \mu m$ increments.

\subsubsection{Carbon and Alumina Samples: Image Acquisition and Analysis}
\label{subsec:carbon_and_alumina_methods}

Images of several porous carbon and alumina samples were taken, over an autocorrelation length (ACL) range of approximately $20 - 80 \: nm$ for the purposes of measuring how the dark-field signal changed as a function of ACL.  This was achieved by taking one set of images with a source-to-grating distance of $L_1 = 20 \: cm$ and grating-to-object distances of $D_{GO} = 10, \: 25, \: 40, \: 55 \: cm$ and another set with $L_1 = 30 \: cm$ and $D_{GO} = 10, \: 20, \: 30, \: 40 \: cm$.  The ACL was calculated for each geometry using a $25 \: keV$ peak energy.  For each acquisition, MPG7 was phase stepped 24 times, at $12 \: \mu m$ increments.

The carbon samples were OMC-6-600, Nuchar, and Calgon-PCB. The alumina samples were ASM-385, SAS-90, and a silica-alumina porosimeter reference standard.  Each carbon sample was in powder form, so images were acquired with each sample in a plastic capillary tube.  ASM-385 and SAS-90 were in the form of compact spheres larger than a capillary tube, so they were placed in a larger plastic tube for imaging.  The silica-alumina was in the form of tiny, approximately cylindrical particles and was also placed into a larger plastic tube.  Because of the relative size difference in the size of the cylinders and the tube, some overlap was inevitable over the path of the ray.

Porosimetry data of each sample was acquired using an ASAP 2020 Plus porosimeter.  Nitrogen adsorption-desorption measurements were taken, and the differential pore volume distribution with respect to pore diameter, $dV / dD$, was computed by the Barrett-Joyner-Halenda (BJH) algorithm \cite{bib:Barrett}.  From this, the total pore volume can be calculated by integrating the differential pore volume with respect to pore diameter.  Since our interferometer was only sensitive to pores around the ACL at acquisition, we limited the integration range to only pores between $10 - 120 \: nm$ and called it the partial pore volume.  The range of pore sizes, pore shapes, and polychromatic X-ray energies make the scattering response of our samples difficult to predict quantitatively, but qualitative predictions can be made.  We expect to see higher scattering signals for samples with a higher partial pore volume in the range of $10 - 120 \: nm$, since the interferometer is designed to be sensitive to scattering structures approximately the size of the ACL.
 
\begin{figure}
\begin{subfigure}[t]{0.48\textwidth}
    \centering
    \includegraphics[keepaspectratio = true, width = \textwidth]{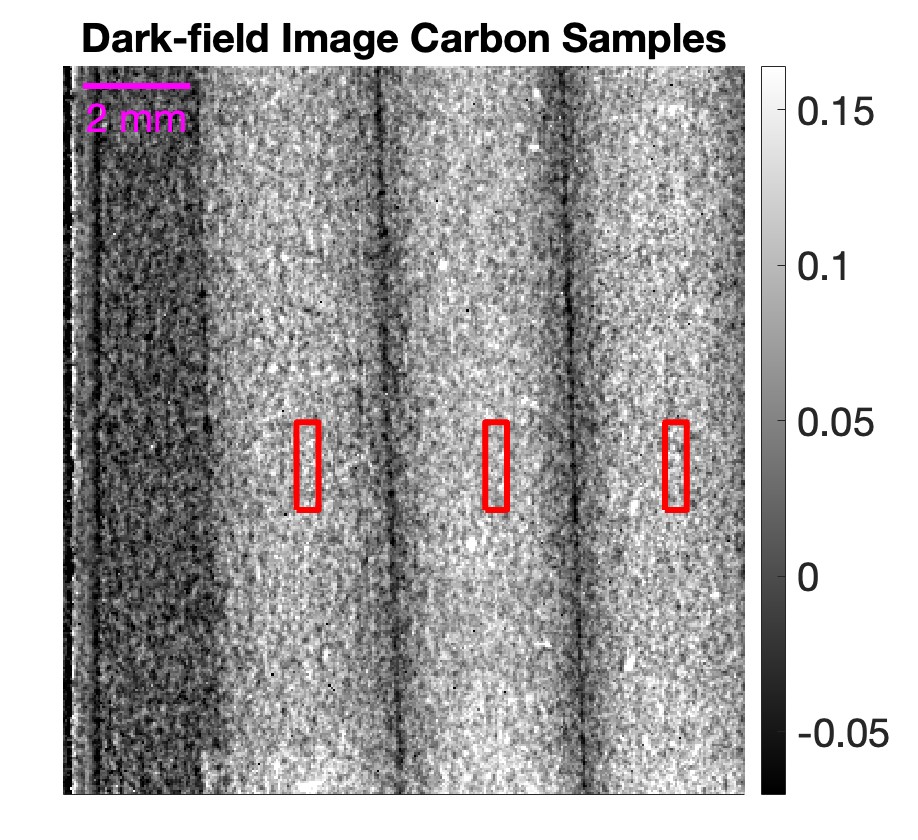}
    \caption{}
    \label{fig:example_carbon_image}
\end{subfigure}
\hfill
\begin{subfigure}[t]{0.48\textwidth}
    \centering
    \includegraphics[keepaspectratio = true, width = \textwidth]{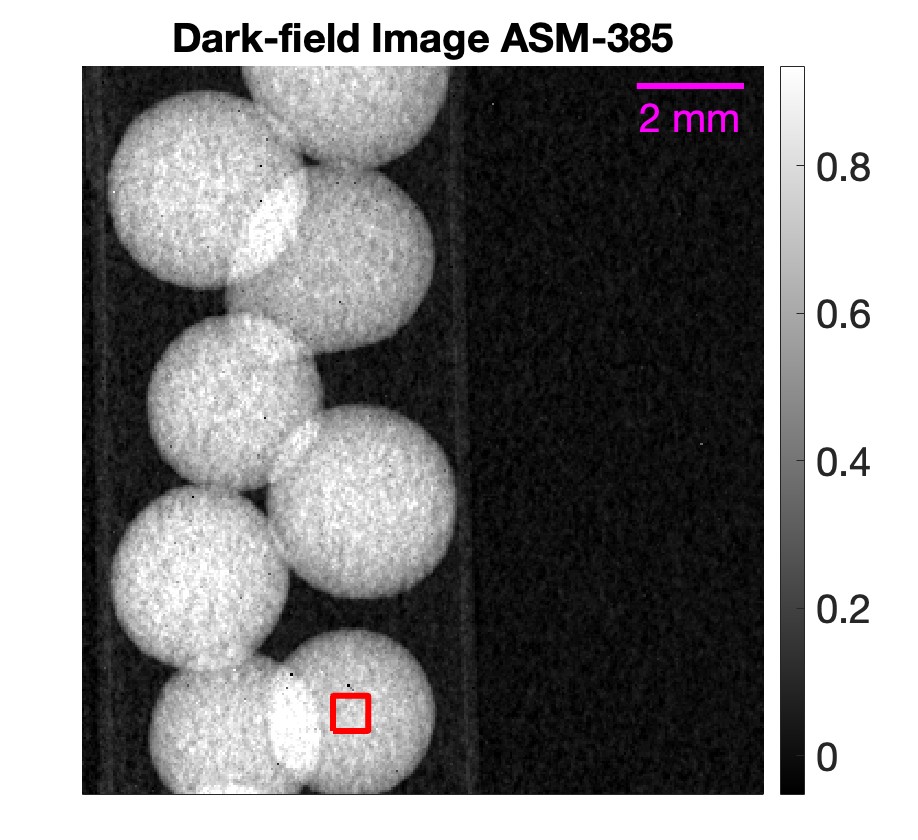}
    \caption{}
    \label{fig:example_alumina_image}
\end{subfigure}
\caption{Example dark-field images with ROIs highlighted of (a) OMC-6-600 (left), Nuchar (middle), Calgon-PCB (right) and (b) ASM-385, acquired with $L_1 = 20 cm$, $D_{GO} = 40 cm$, with $ACL = 37.57 \: nm$}
\label{fig:example_carbon_and_alumina_images}
\end{figure}

Example images of the carbon and alumina samples with their regions of interest (ROI) are shown in Figure \ref{fig:example_carbon_and_alumina_images}, with additional images shown in Results Section \ref{subsec:carbon_and_alumina_results}.  For the carbon samples, a tall rectangular ROI was chosen in the center of each tube, to approximate constant thickness through the path of X-ray.  For the alumina spheres, a small square ROI was chosen at the center of a single sphere, again to approximate constant thickness.  The same sphere was followed for each image, to avoid minor variations between spheres.  For the silica-alumina, overlap was inevitable due to the size of the cylinders relative to their container, but approximately the same ROI was chosen for each geometry.

For each ROI, the attenuation and dark-field signals were averaged to be plotted versus ACL.  Generally speaking, the attenuation signal is expected to not change due to the geometry, but the dark-field signal may change as the ACL is changed, depending on the size and shape of the scattering structures within the sample.  Additionally, the mass attenuation coefficient, $\mu / \rho$, will be approximately constant within each set of carbon and alumina samples.  This is clearly true for the carbon samples, but for the set of alumina samples there is more nuance.  The pure alumina samples (ASM-385 and SAS-90) obviously have the same mass attenuation coefficient, but this is not obvious for the silica-alumina compound.  However, according to NIST's XCOM database, \cite{bib:XCOM}, silica and alumina have approximately the same $\mu / \rho$, meaning the compound also has approximately the same mass attenuation coefficient as the other alumina samples.

However, their mass thickness, $\rho t$, varies due to the differences in the packing density (for the carbon samples), the porosity differences between the samples, and the pixel-to-pixel thickness variation (however small).  This can be corrected if we normalize the log-scale dark-field image by the log-scale attenuation image. Assuming the mass attenuation coefficient, $\mu / \rho$, is approximately constant over the projection of one pixel, the attenuation contrast can be represented as

\begin{equation}
    -\ln \left(\frac{a_{0,sample}}{a_{0,blank}} \right) = \frac{\mu}{\rho} \times \rho t
\end{equation}

The dark-field contrast can be represented in a similar manner, using the linear diffusion coefficient, $\epsilon / \rho$, which depends on the pore microstructures within the sample:

\begin{equation}
    -\ln \left( \frac{V_{sample}}{V_{blank}} \right) = \frac{\epsilon}{\rho} \times \rho t
\end{equation}

Thus, normalizing the dark-field signal by the attenuation signal corrects for the mass thickness, $\rho t$, and all left is the linear diffusion coefficient scaled by the mass attenuation coefficient.  Since the mass attenuation coefficient is expected to be the same for the three carbon samples and the same for the three alumina samples, we expect the average normalized dark-field signal to follow the trend of the partial pore volume, even when the dark-field contrast curve without normalization may not.

\section{Results}
 
\subsection{Theory Results}
\label{subsec:theory_results}

The theory presented in Section \ref{subsec:theory} is evaluated in several ways.  The intensity profile is calculated for the 2D MPG, and the 2D MPG under the $l = 0$ approximation is compared to the case with sufficient detector blur in the y-direction and a sufficiently small $p_y$.  The $l = 0$ approximation is then compared to the 1D MPG under the conditions of $\alpha_x = \alpha/2$, $p_x = 2p$, $\alpha_y = 1$, which corresponds to the removal of the staggering of the grating bars.  Lastly, the visibility as a function of grating-to-detector distance, $z$, is calculated for an \textit{ideal} MPG with $(\pi, 0)$ phase heights and an MPG with phase heights that match MPG7 and MPG8, $(\pi/2, \pi/8)$.  The same setup conditions from our experiments were used in our simulations.

\subsubsection{The Validity of the $l = 0$ Approximation}

\begin{figure}[H]
\begin{subfigure}[t]{0.48\textwidth}
    \centering
    \includegraphics[keepaspectratio = true, width = \textwidth]{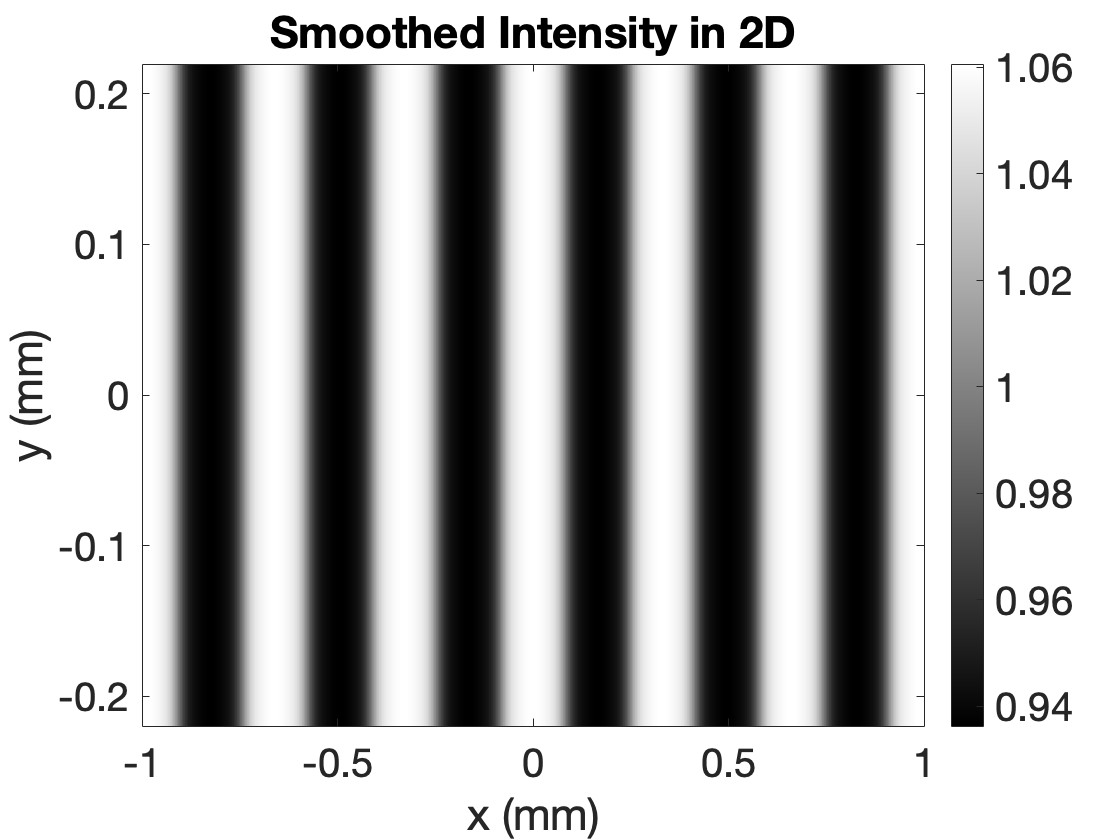}
    \subcaption{}
    \label{fig:2D_intensity}
\end{subfigure}
\hfill
\begin{subfigure}[t]{0.48\textwidth}
    \centering
    \includegraphics[keepaspectratio = true, width = \textwidth]{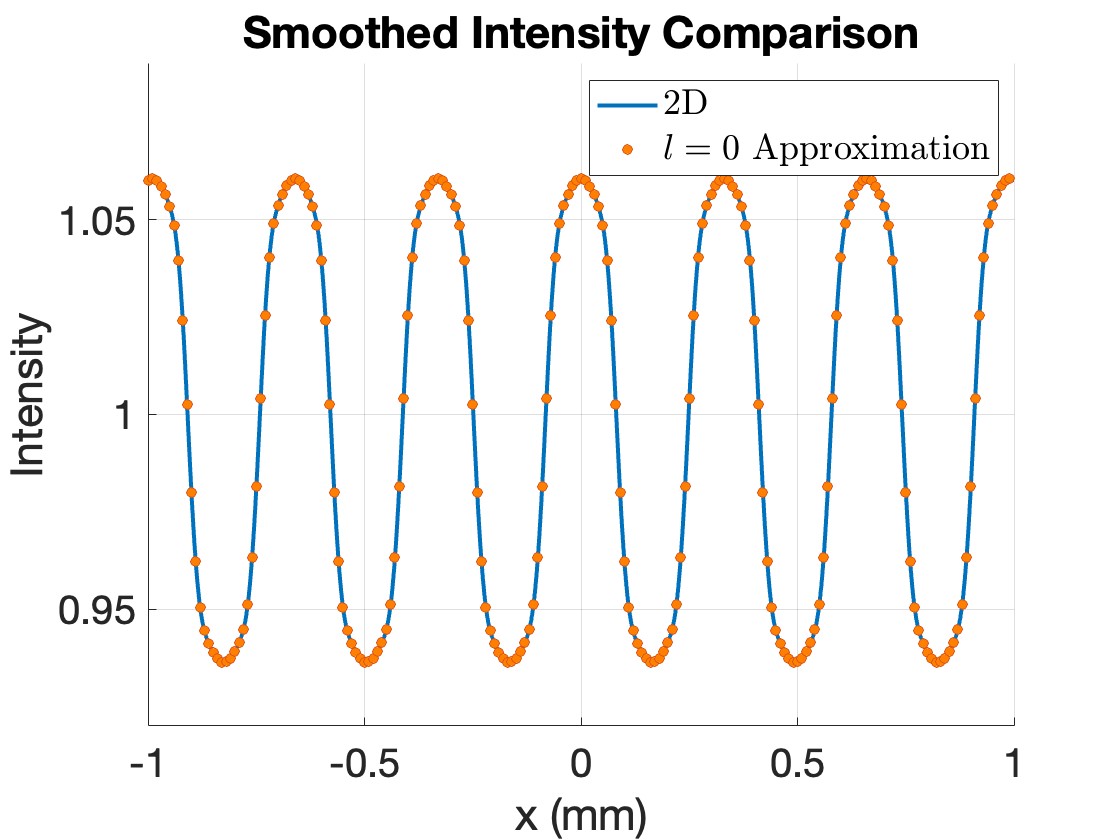}
    \caption{}
    \label{fig:2Dapprox_verification}
\end{subfigure}
\caption{Verification of the $l = 0$ approximation using a RectMPG (Table \ref{tab:MPG_parameters}) at $25 \: keV$, $L = 110 \: cm$, $z = 70 \: cm$.  (a) Theoretical 2D intensity calculation after smoothing (b) Overlay of a single line through the smoothed 2D intensity profile and the $l = 0$ Approximation.}
\end{figure}

The $l = 0$ approximation is validated by showing that the analytical 2D field calculations are well approximated by the $l = 0$ approximation when the blur in the y-dimension is sufficiently high.  The 2D field amplitude at the detector is found from Equation \ref{eq:2D_fieldAmplitude}.  The intensity is then calculated as $I = |U(x,y,z|^2$.  A monochromatic case of $25 \: keV$ is simulated for a realistic MPG with a rectangular envelope with a source-to-detector distance of $L = 110 \: cm$ and grating-to-detector distance of $z = 70 \: cm$.  \textcolor{black}{A realistic Gaussian source with $\sigma = 3.14 \: \mu m$ and generalized Gaussian PSF with $\sigma = 57.87 \: \mu m$ and $k = 0.9$ were used.  These parameters were determined experimentally for the Keck imaging system, as explained in Section \ref{subsec:experimental_methods}, leading to Figure \ref{fig:2D_intensity}.}  Under the same conditions, the intensity using the $l = 0$ approximation from Equation \ref{eq:1D_approx_intensity} is calculated and is overlaid in Figure \ref{fig:2Dapprox_verification}.  There is excellent agreement between the blurred fully calculated 2D intensity and the intensity calculated using the $l = 0$ approximation, verifying that this approximation is appropriate for realistic MPG and setup conditions since the source and detector blur in the y-dimension is sufficiently high enough to remove the other harmonics. The $l = 0$ approximation required about 3000 times less computations for a sample rate of $0.1 \: \mu m$ and a vertical field-of-view (FOV) of $0.3 \: mm$.  This FOV was necessary for the convolution of the Gaussian source (magnified) and detector PSF over 3 standard deviations.

\begin{figure}[H]
\begin{subfigure}[t]{0.48\textwidth}
    \centering
    \includegraphics[keepaspectratio = true, width = \textwidth]{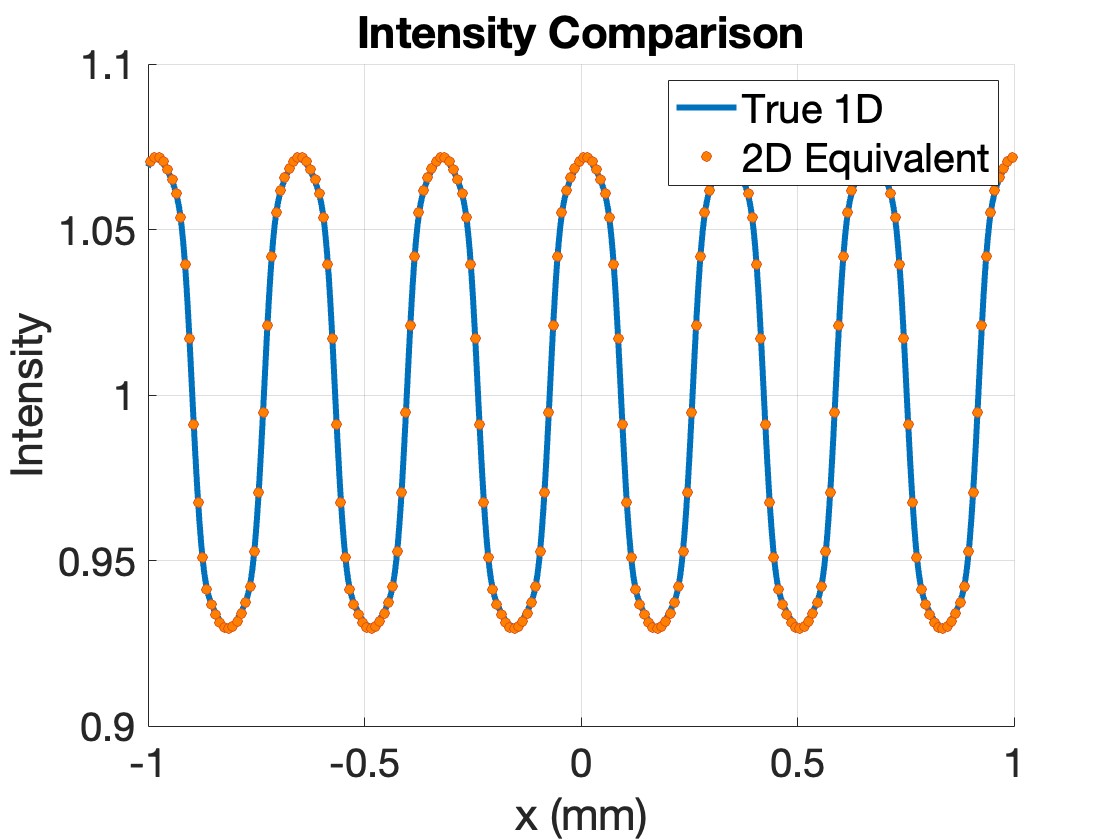}
    \caption{}
    \label{fig:1D_and_2D_equivalent_intensity_profile}
\end{subfigure}
\hfill
\begin{subfigure}[t]{0.48\textwidth}
    \centering
    \includegraphics[keepaspectratio = true, width = \textwidth]{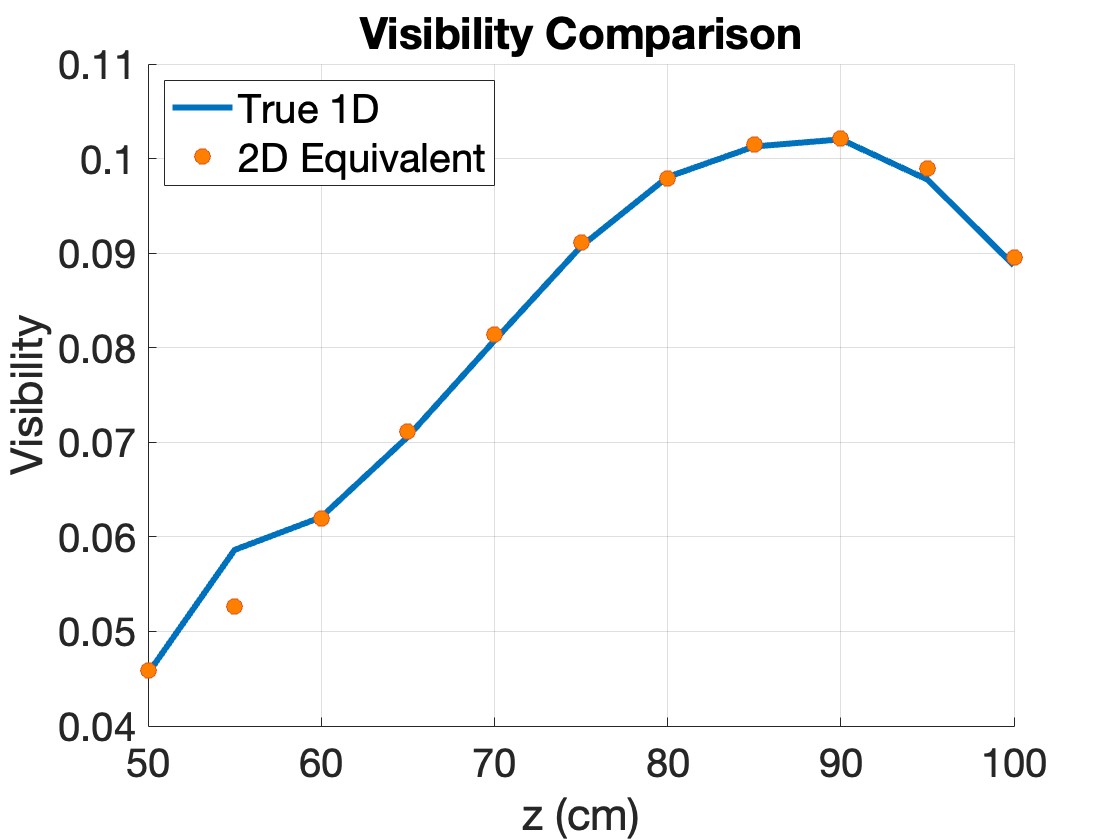}
    \caption{}
    \label{fig:1D_and_2D_equivalent_visibility_comparison}
\end{subfigure}
\caption{Verification of the $l = 0$ approximation with the equivalent 1D RectMPG with parameters $\alpha = 0.5$, $p = 1 \: \mu m$, $W = 120 \: \mu m$, $h_2 = \pi / 2$, $h_1 = \pi / 8$, $E_D = 25 \: keV$, made of Gold. Calculated using a monochromatic $25 \: keV$ source, a fixed source-to-detector distance of $L = 110 \: cm$, and realistic source and detector blur (a) Overlay of the 1D intensity with the 2D equivalent found using the $l = 0$ approximation (b) Visibility versus grating-to-detector distance comparison between the true 1D and 2D equivalent found using the $l = 0$ approximation.}
\label{fig:1D_and_2D_equivalent_comparison}
\end{figure}

Furthermore, as seen in Figure \ref{fig:1D_and_2D_equivalent_comparison}, our methodology is shown to be consistent with the 1D MPG from previous work, \cite{bib:HidrovoMeyerRSI}, when $\alpha_y = 1$, $p_x = 2p$ and $\alpha_x = \alpha / 2$ where the 1D pitch is $p$ and $\alpha$ is the grating's duty cycle.  Over a range of geometries, the fringe profile of the true 1D MPG was calculated using Equation \ref{eq:new_1D_fieldAmplitude}, and the fringe profile of the equivalent 2D MPG was calculated using the $l = 0$ approximation, Equation \ref{eq:1D_approx_intensity}.  The visibility was calculated as a function of grating-to-detector distance, $z$, for a fixed source-to-detector distance, $L$, and good agreement was shown between the two methods.  It is evident that the 2D MPG is reduced to the 1D MPG when the staggering of the grating bars is removed when $\alpha_y = 1$ and by scaling $p_x$ and $\alpha_x$ to the equivalent 1D MPG parameters.

\subsubsection{Theoretical Visibility}
\label{subsec:theoretical_visibility}

Simulations using the theory presented in Section \ref{subsec:theory} were performed for the case that matched the RectMPG parameters of MPG7 and MPG8 with heights $(h_2, h_1)$ (Figure \ref{fig:MPG_schematic}) tuned to a phase shift of $(\pi/2, \pi/8)$ at $25 \: keV$. Other simulation parameters matched our experimental setup, with a fixed source-to-detector distance of $L = 110 \: cm$ and a range of grating-to-detector distances of $z = 56-98 \: cm$.  We used a $45 \: kVp$ tungsten anode source, with a $0.2 \: mm$ beryllium filter, calculated using the TASMIP spectra calculator, \cite{bib:TASMIP}.  \textcolor{black}{A Gaussian source of $3.14 \: \mu m$ and generalized Gaussian PSF with $\sigma = 57.87 \: \mu m$ were used.  These parameters were determined experimentally, as explained in Section \ref{subsec:experimental_methods}.}

\begin{figure}
\begin{subfigure}[b]{0.48\textwidth}
    \centering
    \includegraphics[keepaspectratio = true, width = \textwidth]{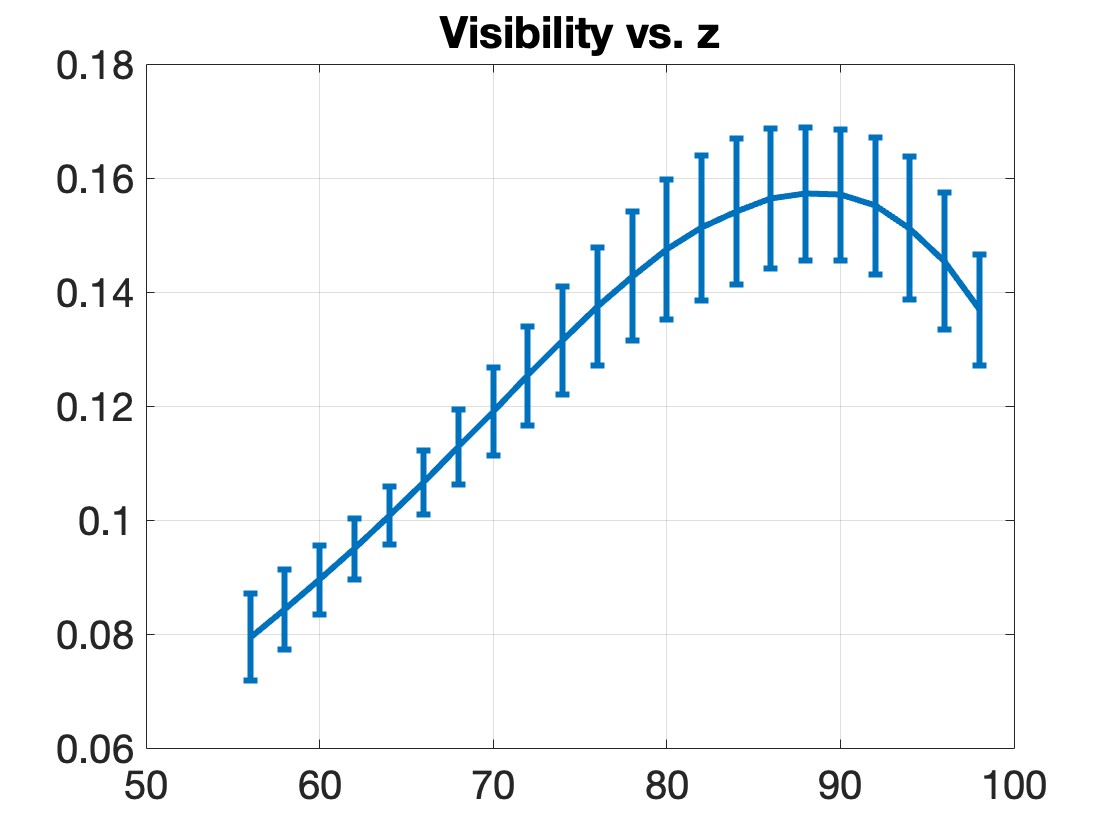}
    \caption{}
    \label{fig:theory_vis_vs_z_idealMPG}
\end{subfigure}
\begin{subfigure}[b]{0.48\textwidth}
    \centering
    \includegraphics[keepaspectratio = true, width = \textwidth]{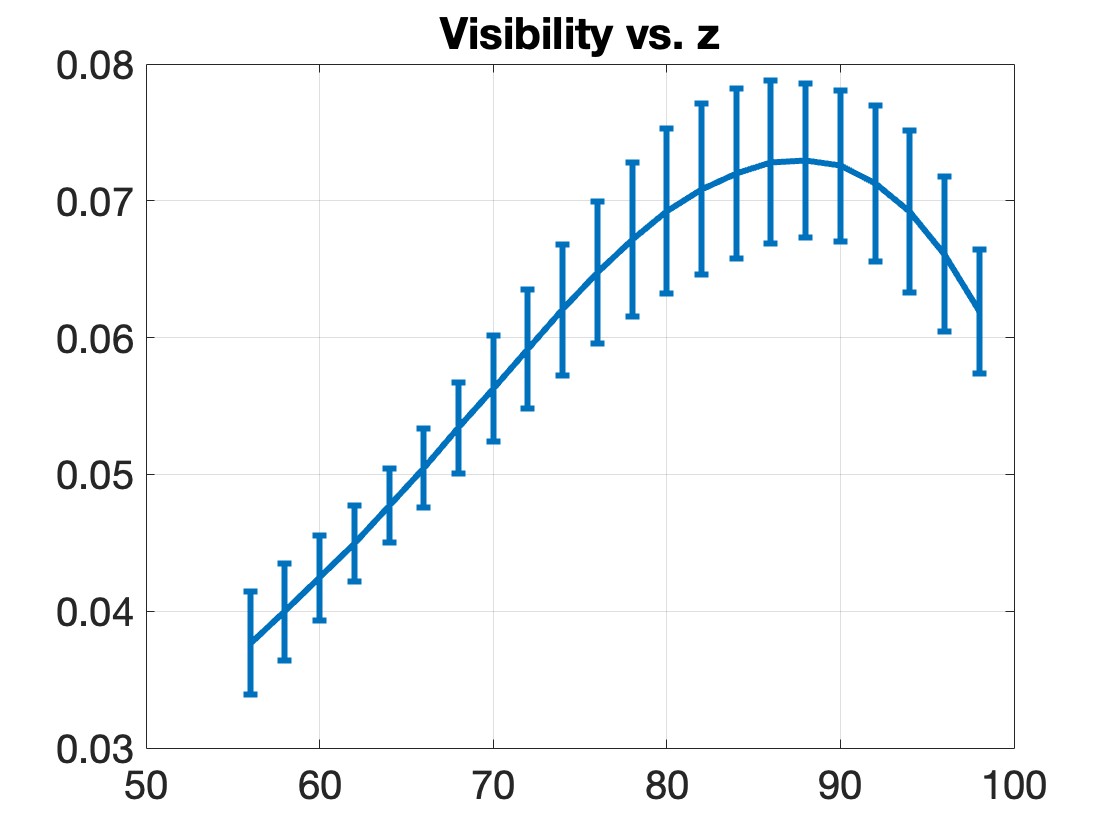}
    \caption{}
    \label{fig:theory_vis_vs_z_experimentalMPG}
\end{subfigure}
\caption{Theoretical Visibility vs. grating-to-detector distance, $z$, for (a) an ideal $(\pi, 0)$ RectMPG and (b) a $(\pi/2, \pi/8)$ RectMPG, as listed in Table \ref{tab:MPG_parameters}.  All parameters are the same between the two simulations except for the phase heights of the grating envelope function, $h_2$ and $h_1$.}
\label{fig:theory_vis_vs_z}
\end{figure}

We've previously shown the best visibility for a RectMPG is when the height $h_1=0$ and $h_2$ corresponds to a $\pi$ shift at the design energy for the grating material (in our case, Gold at $25 \: keV$) \cite{bib:HidrovoMeyerRSI}. To test this ideal case, theoretical simulations were also performed for the case of an ideal RectMPG, with $(\pi, 0)$ phase heights. The results are compared with the case of the $(\pi/2, \pi/8)$ RectMPG in Figure \ref{fig:theory_vis_vs_z}.  The shape of the curves is approximately the same, with the primary difference being the visibility of the $(\pi/2, \pi/8)$ grating is lower than the $(\pi, 0)$ grating, highlighting the importance of maximizing the difference in the phase heights for a RectMPG.  However, this is limited by cost and fabrication stability.  The error bars represent the standard deviation in the visibility, which results from the oscillations in the visibility on a pixel-to-pixel basis due to a finite number of phase steps.

\subsection{Experiment Results}

\subsubsection{Experimental Visibility Compared With Theory}

\begin{figure}
    \centering
    \includegraphics[keepaspectratio=true,width = 0.7\textwidth]{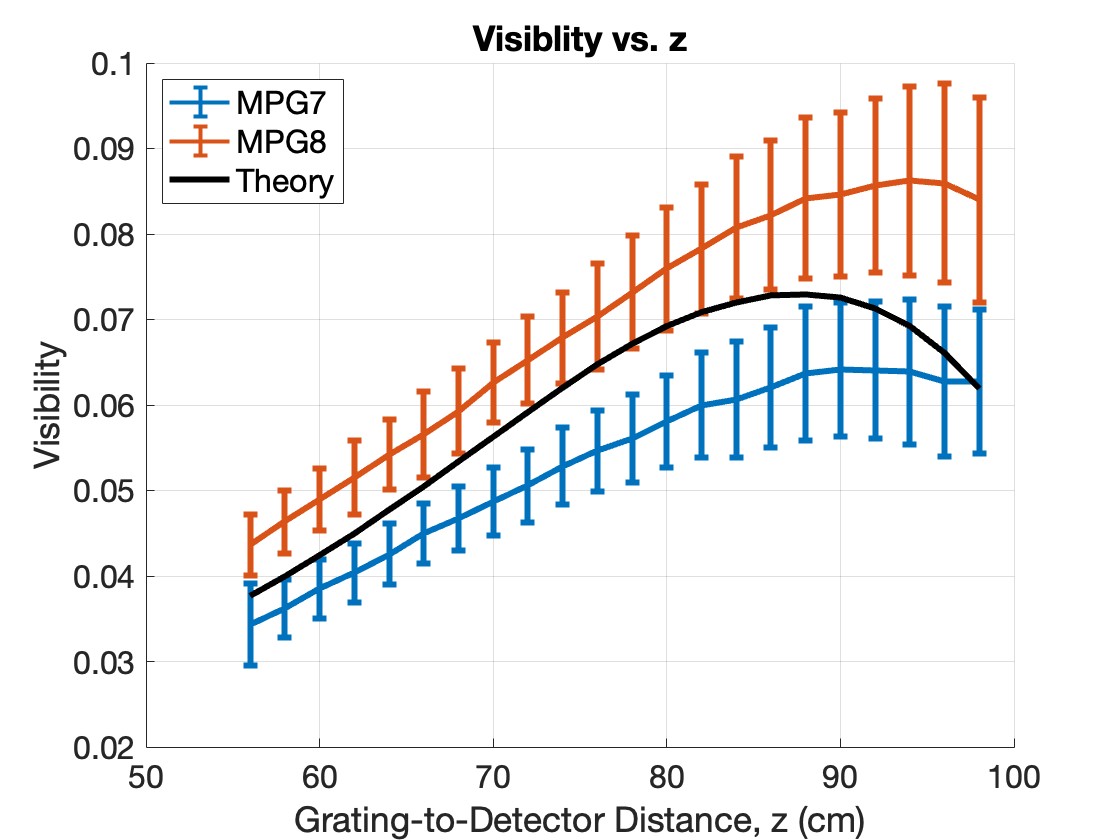}
    \caption{\textcolor{black}{Comparison of Visibility vs. Grating-to-Detector Distance, $z$, for MPG7, MPG8, and the presented theory.  The visibility was measured as as function of grating-to-detector distance, $z$ for a fixed source-to-detector distance, $L = 110 \: cm$, with details in Section \ref{subsec:exper_vis_measurements}.}}
    \label{fig:vis_vs_z_overlay_MPG7_MPG8_theory_genGaussian_PSF}
\end{figure}

\textcolor{black}{The visibility produced by MPG7 and MPG8 was experimentally measured as a function of grating-to-detector distance, $z$, for a fixed source-to-detector distance, $L = 110 \: cm$, as described in Section \ref{subsec:exper_vis_measurements}.  The theoretical visibility, as described in Section \ref{subsec:theoretical_visibility} and shown in Figure \ref{fig:theory_vis_vs_z_experimentalMPG}, was directly compared with the experimentally obtained MPG7 and MPG8 visibility curves, as shown in Figure \ref{fig:vis_vs_z_overlay_MPG7_MPG8_theory_genGaussian_PSF}.  It is seen that the theoretical visibility lies between the measured visibility of each grating.  We suspect the differences seen in the visibility produced by each grating results from variabilities in the height of the grating structures for MPG7 and MPG8, as shown in Table \ref{tab:MPG_height_measurements}.  It is seen that MPG8 has a larger difference in the height of the grating structures, $\Delta h$, leading to a higher visibility.}

\subsubsection{Carbon and Alumina Samples: Analysis Results}
\label{subsec:carbon_and_alumina_results}

\begin{figure}
\begin{subfigure}[t]{0.32\textwidth}
    \centering
    \includegraphics[keepaspectratio = true, width = \textwidth]{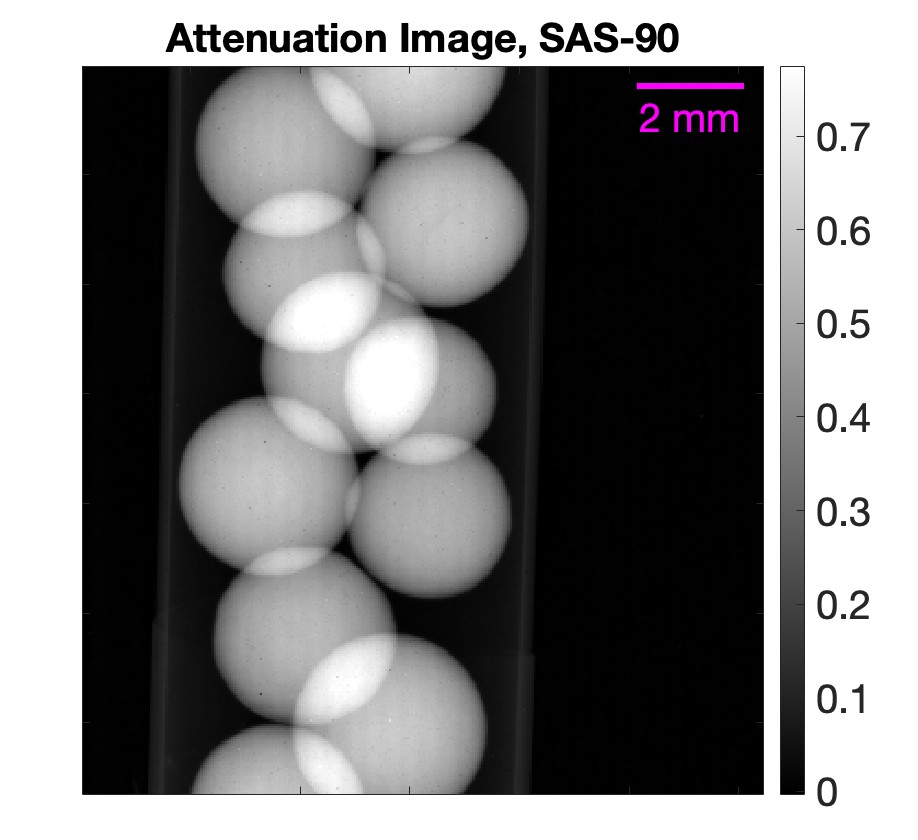}
    \caption{}
    \label{fig:alumina_attenuation}
\end{subfigure}
\hfill
\begin{subfigure}[t]{0.32\textwidth}
    \centering
    \includegraphics[keepaspectratio = true, width = \textwidth]{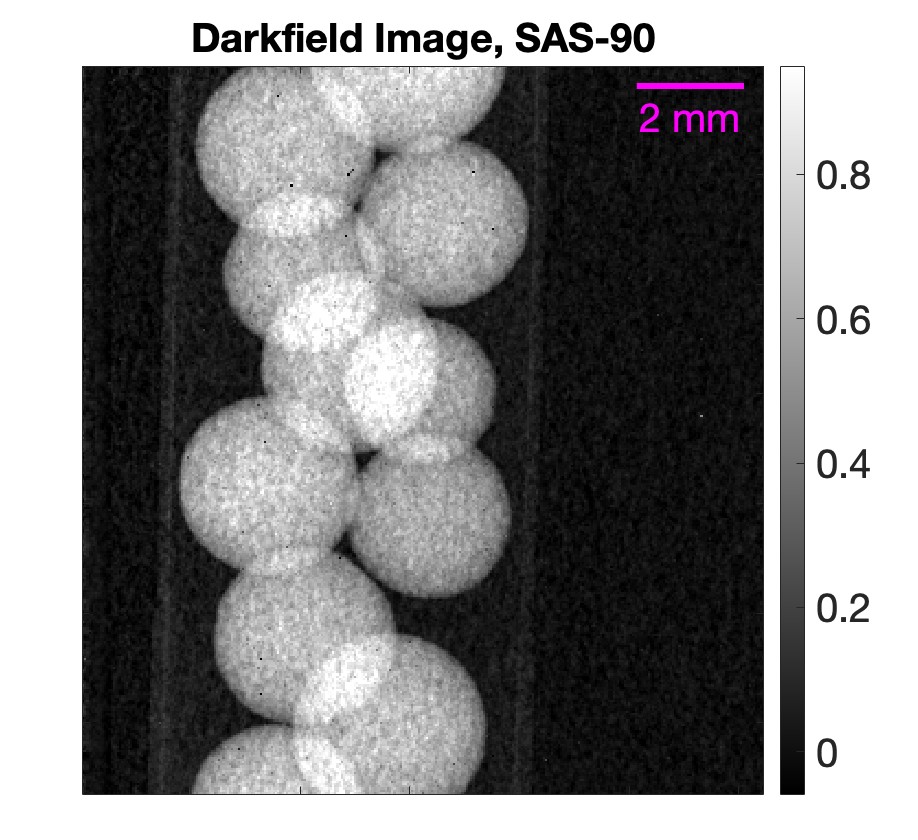}
    \caption{}
    \label{fig:alumina_attenuation}
\end{subfigure}
\hfill
\begin{subfigure}[t]{0.32\textwidth}
    \centering
    \includegraphics[keepaspectratio = true, width = \textwidth]{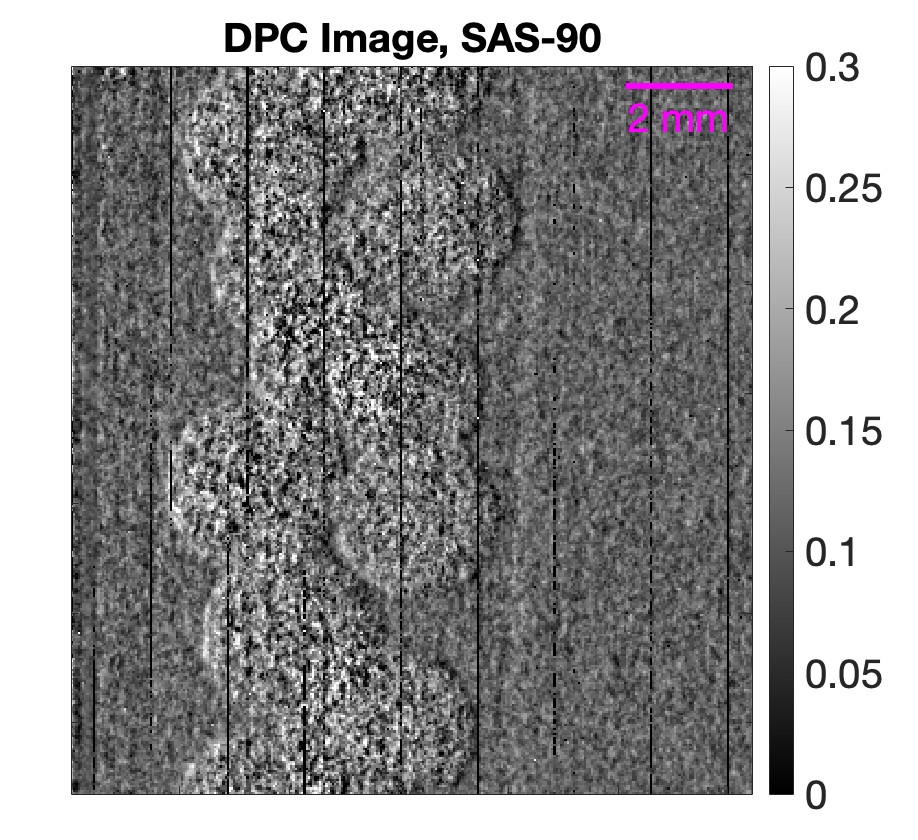}
    \caption{}
    \label{fig:alumina_attenuation}
\end{subfigure}
\caption{Attenuation (a), dark-field (b), and differential-phase contrast (c) images of SAS-90, taken at a source-to-detector distance of $L = 110 \: cm$, the source-to-grating distance of $L_1 = 20 \: cm$, and grating-to-object distance of $D_{GO} = 40 \: cm$.}
\label{fig:additional_alumina_images_all_three}
\end{figure}

Attenuation, dark-field, and differential-phase contrast images of several carbon and alumina samples were acquired using the methods described in Section \ref{subsec:carbon_and_alumina_methods}.  Figure \ref{fig:additional_alumina_images_all_three} shows examples of the SAS-90 alumina sample.  The DPC images are visible for the ASM-385 and SAS-90 alumina samples, where the spheres' characteristic bright and dark sides are seen.  The other samples did not have a significant DPC signal.

\begin{table}[]
    \centering
    \begin{tabular}{|c||c|c|}
        \hline
         \textbf{Sample} & \textbf{Total Pore Volume (cc/g)} & \textbf{Partial Pore Volume (cc/g)} \\
         \hline
         OMC-6-600 & 0.464 & 0.0168 \\
         \hline
         Nuchar & 0.352 & 0.104 \\
         \hline
         Calgon-PCB & 0.0899 & 0.00711 \\
         \hhline{|=||=|=|}
         ASM-385 & 0.474 & 0.445 \\
         \hline
         SAS-90 & 0.408 & 0.334 \\
         \hline
         Silica-Alumina & 0.6699 & 0.220 \\
         \hline
    \end{tabular}
    \caption{Total Pore Volume and Partial Pore Volume within the range of $10-120 \: nm$ for each carbon and alumina sample.}
    \label{tab:pore_volume_measurements}
\end{table}

\begin{figure}
\begin{subfigure}[t]{0.48\textwidth}
    \centering
    \includegraphics[keepaspectratio = true, width = \textwidth]{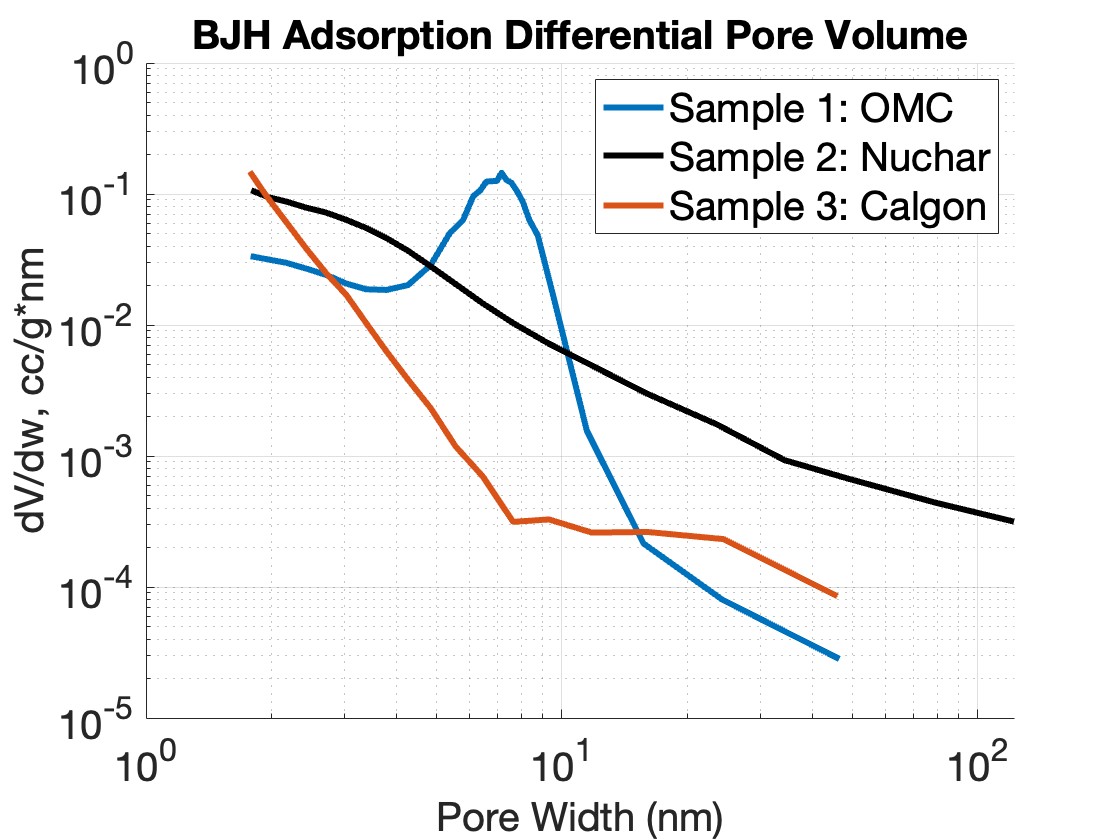}
    \caption{}
    \label{fig:carbon_dVdw}
\end{subfigure}
\hfill
\begin{subfigure}[t]{0.48\textwidth}
    \centering
    \includegraphics[keepaspectratio = true, width = \textwidth]{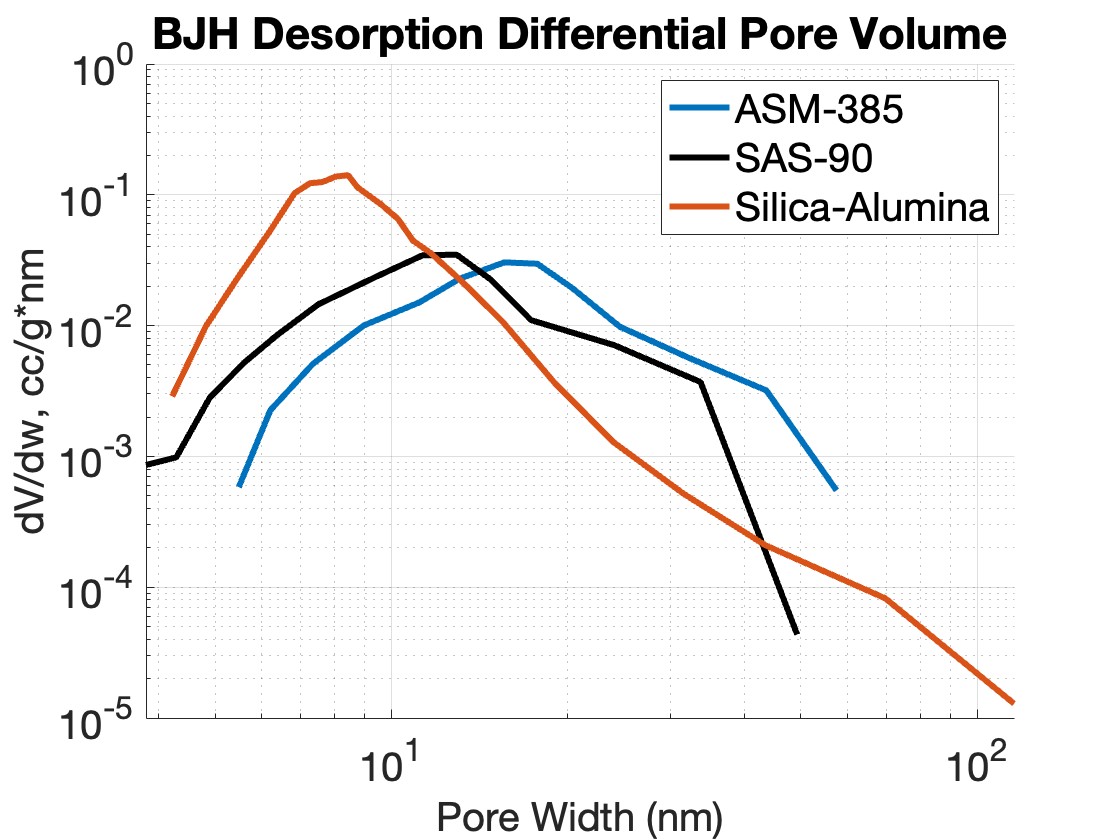}
    \caption{}
    \label{fig:alumina_dVdw}
\end{subfigure}
\caption{Differential Pore Volume Measurements of the (a) carbon and (b) alumina samples.  The partial pore volume is calculated by integrating from $10 - 120 \: nm$.}
\label{fig:dVdw_measurements}
\end{figure}

The mean normalized dark-field signal---the dark-field image divided by the attenuation image---was measured for each carbon and alumina sample, to be qualitatively compared with the partial pore volume of the samples, with details in Section \ref{subsec:carbon_and_alumina_methods}.  For the purposes of calculating the partial pore volume, BJH adsorption measurements were used for the carbon samples while BJH desorption was used for the alumina samples, due to the differences in pore \textit{shapes} commonly seen between these samples \cite{bib:Kruk, bib:Gregg}.  Figure \ref{fig:dVdw_measurements} shows the differential pore volume measurements.  Integration was performed from $10 - 120 \: nm$ (or the maximum pore size measured) to calculate the partial pore volume.

Table \ref{tab:pore_volume_measurements} shows the total and partial pore volume measurements.  For the carbon samples, it is seen that Nuchar has the highest partial pore volume within the range of $10 - 120 \: nm$ and OMC-6-600 and Calgon-PCB are both lower by about an order of magnitude.  ASM-385 has a higher partial pore volume for the alumina samples than SAS-90, and the silica-alumina sample has the lowest partial pore volume.

\begin{figure}
    \centering
    \includegraphics[keepaspectratio = true, width = \textwidth]{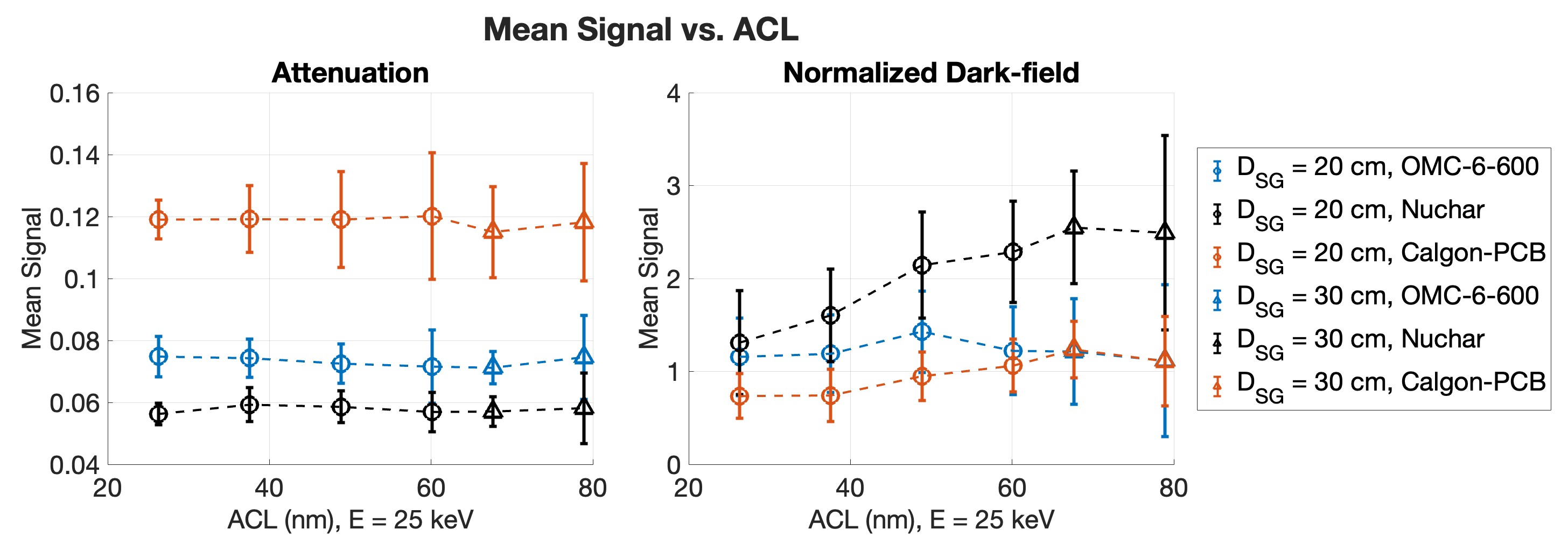}
    \caption{Mean attenuation and mean normalized dark-field signal as a function of ACL for the carbon samples.   The ACL was calculated using $E = 25 \: keV$.}
    \label{fig:mean_norm_darkfield_carbon}
\end{figure}

\begin{figure}
    \centering
    \includegraphics[keepaspectratio = true, width = \textwidth]{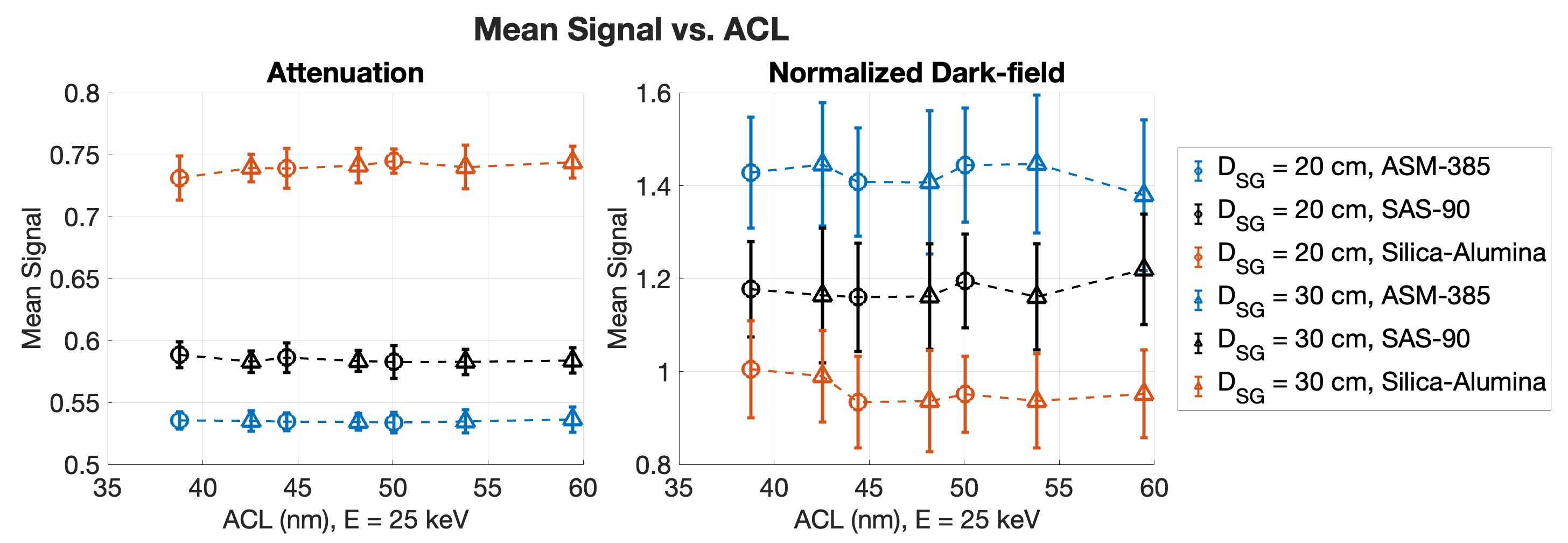}
    \caption{Mean attenuation and mean normalized dark-field signal as a function of ACL for the alumina samples.  The ACL was calculated using $E = 25 \: keV$.}
    \label{fig:mean_norm_darkfield_alumina}
\end{figure}

The same trends are seen when looking at the normalized dark-field signal versus ACL curves in Figures \ref{fig:mean_norm_darkfield_carbon} and \ref{fig:mean_norm_darkfield_alumina}.  For the carbon samples, the normalized dark-field signal is highest for Nuchar, which had the highest partial pore volume.  The OMC-6-600 and Calgon-PCB samples both have lower mean signal, with OMC-6-600 being higher on the lower end of the ACL range.  This is likely due to the partial pore volume being slightly higher in the OMC-6-600, especially if the pore width range of interest was expanded to include the peak differential pore volume.  The mean signal of the Calgon-PCB sample is about the same as the OMC-6-600 sample at the higher end of the ACL range, which may be due to the differential pore volume for Calgon-PCB being larger at the higher end of the pore width range, as seen in Figure \ref{fig:carbon_dVdw}. For the alumina samples, the normalized dark-field signal follows the trend of the partial pore volume, with the ASM-385 sample having the highest signal, followed by SAS-90, then the silica-alumina sample.

\subsubsection{Anchovy Images}
\label{subsec:anchovy_images}

\begin{figure}
\begin{subfigure}[t]{0.32\textwidth}
    \centering
    \includegraphics[keepaspectratio = true, width = \textwidth]{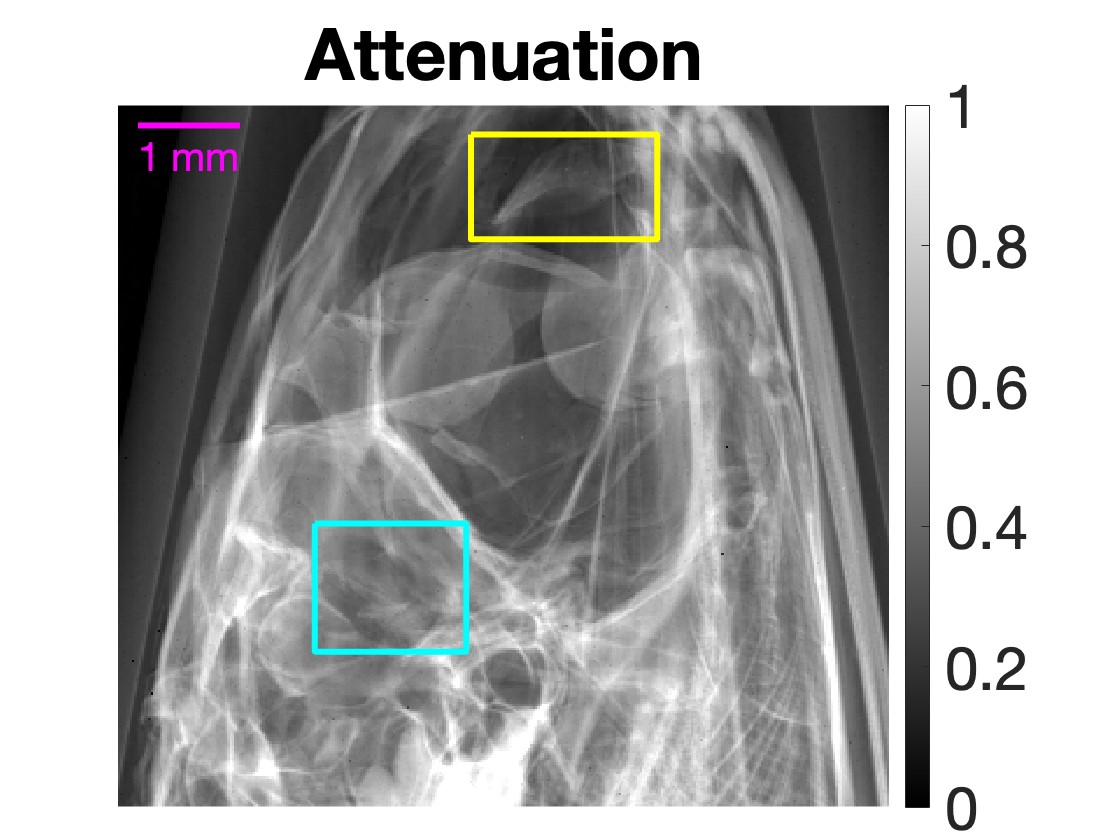}
    \caption{}
    \label{fig:anchovy_attenuation_MPG7}
\end{subfigure}
\hfill
\begin{subfigure}[t]{0.32\textwidth}
    \centering
    \includegraphics[keepaspectratio = true, width = \textwidth]{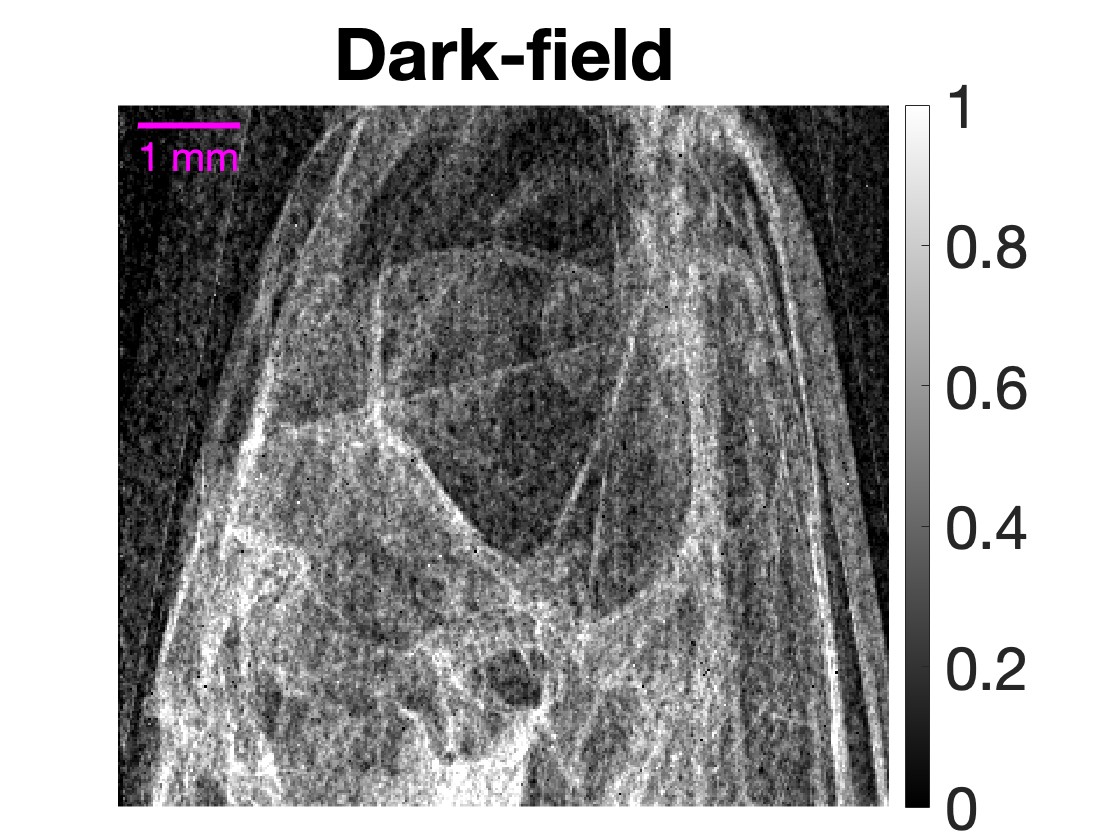}
    \caption{}
    \label{fig:anchovy_darkfield_MPG7}
\end{subfigure}
\hfill
\begin{subfigure}[t]{0.32\textwidth}
    \centering
    \includegraphics[keepaspectratio = true, width = \textwidth]{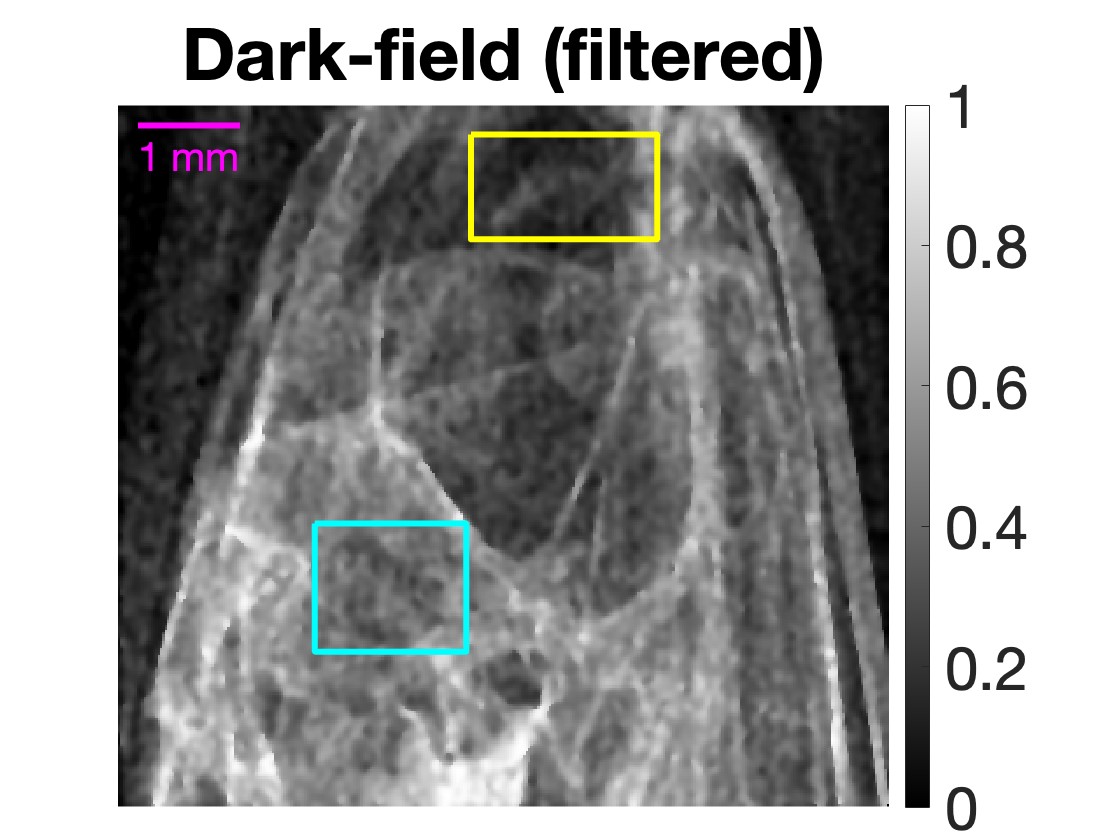}
    \caption{}
    \label{fig:anchovy_darkfield_filtered_MPG7}
\end{subfigure}

\centering
\hfill
\begin{subfigure}[t]{0.24\textwidth}
    \centering
    \includegraphics[keepaspectratio = true, width = \textwidth]{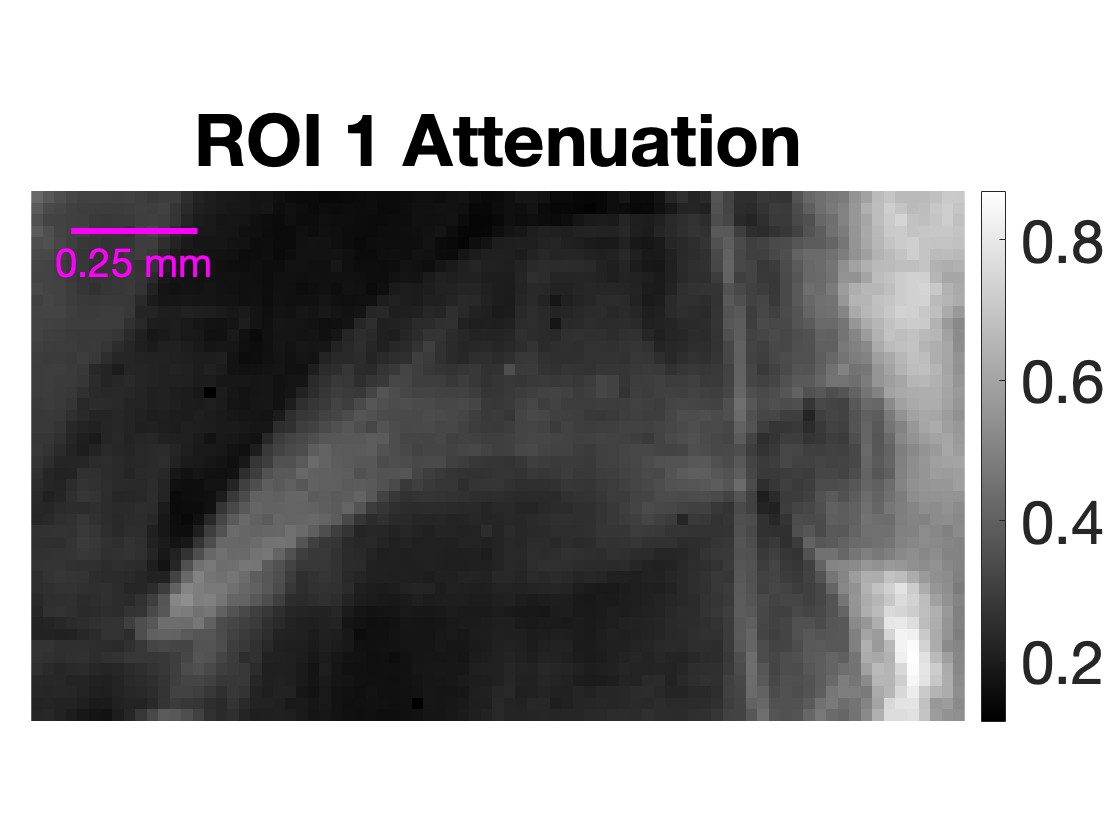}
    \caption{}
    \label{fig:anchovy_attenuation_zoomed_MPG7}
\end{subfigure}
\hfill
\begin{subfigure}[t]{0.24\textwidth}
    \centering
    \includegraphics[keepaspectratio = true, width = \textwidth]{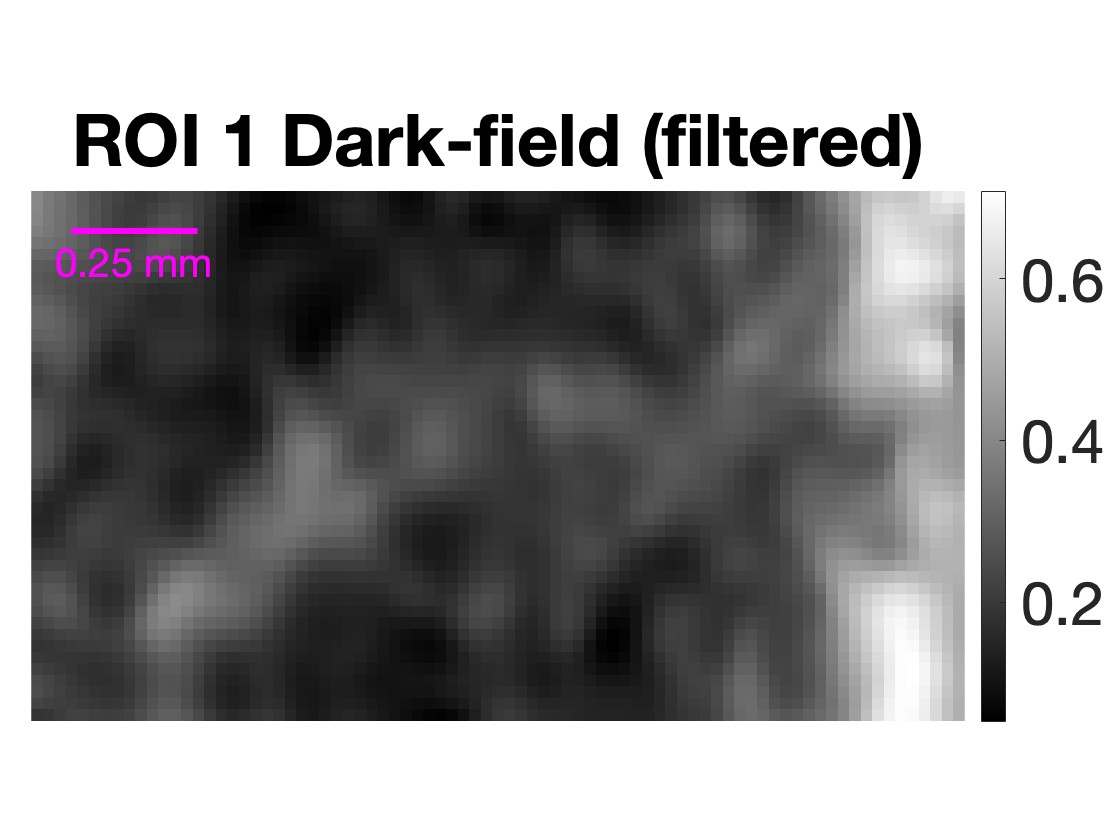}
    \caption{}
    \label{fig:anchovy_darkfield_filtered_zoomed_MPG7}
\end{subfigure}
\hfill
\begin{subfigure}[t]{0.24\textwidth}
    \centering
    \includegraphics[keepaspectratio = true, width = \textwidth]{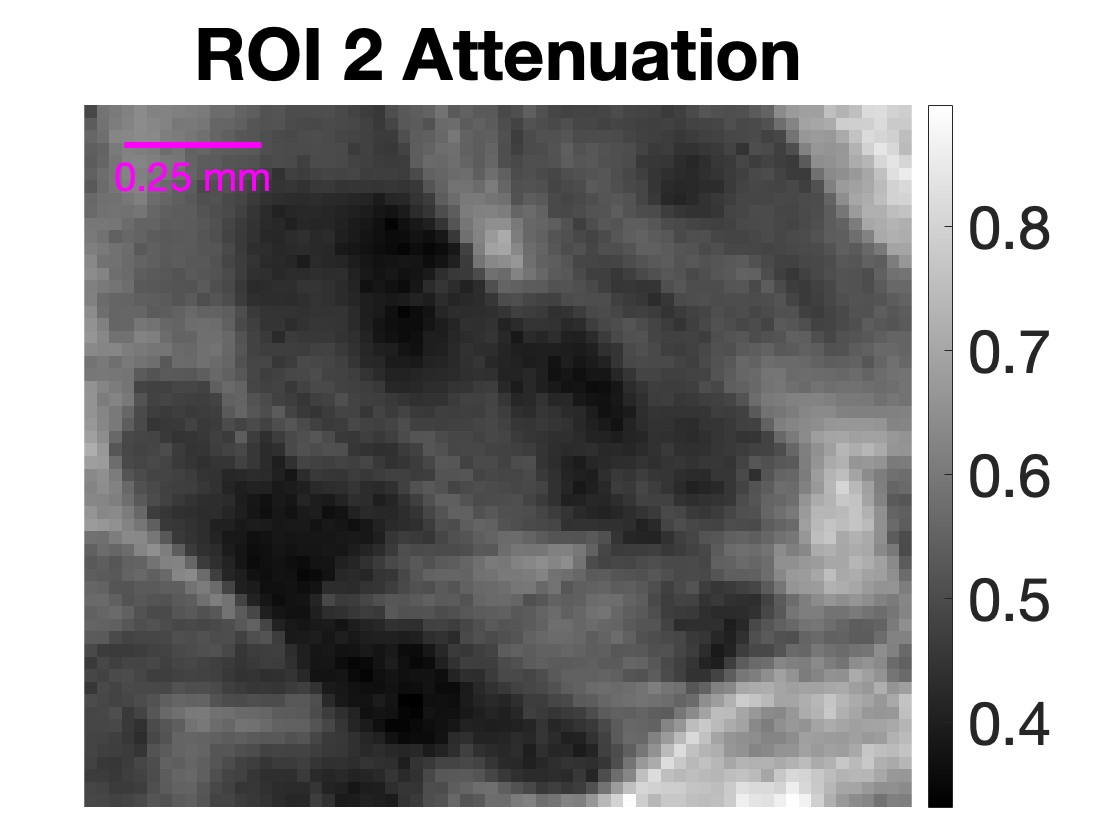}
    \caption{}
    \label{fig:anchovy_attenuation_zoomed_MPG7}
\end{subfigure}
\hfill
\begin{subfigure}[t]{0.24\textwidth}
    \centering
    \includegraphics[keepaspectratio = true, width = \textwidth]{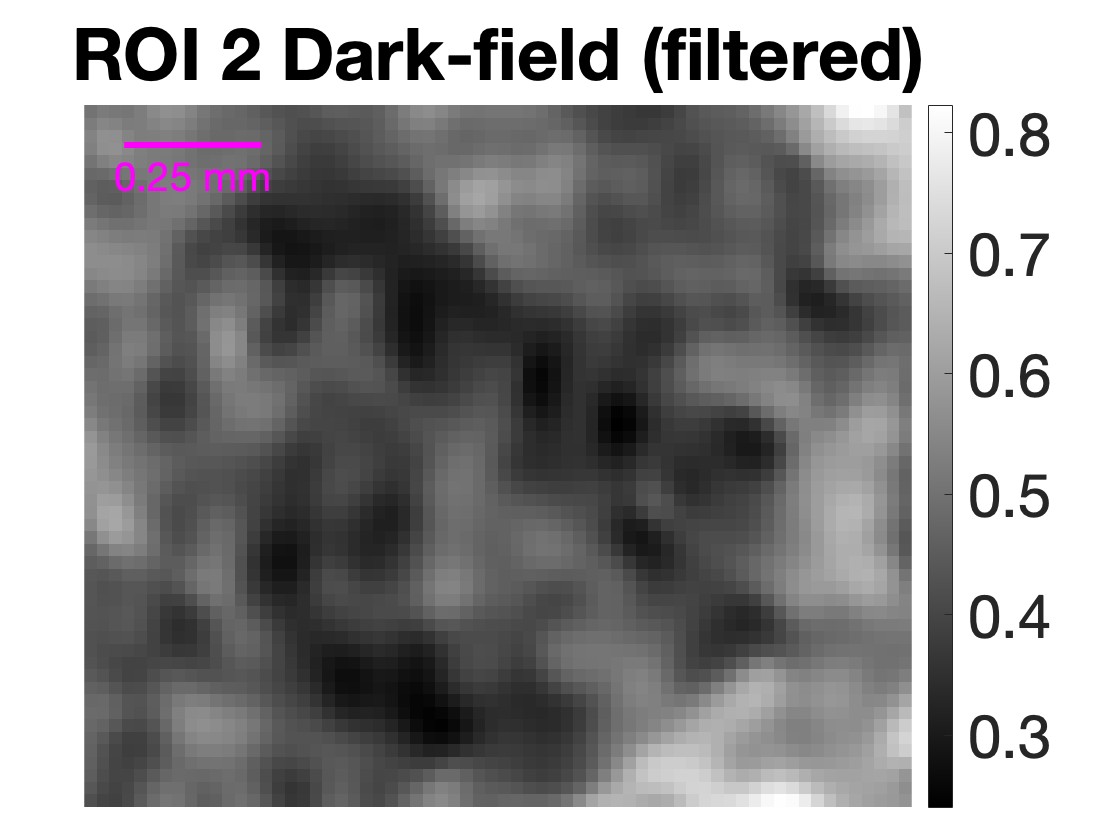}
    \caption{}
    \label{fig:anchovy_darkfield_filtered_zoomed_MPG7}
\end{subfigure}
\hspace*{\fill}

\caption{Anchovy images acquired using MPG7.  Images were acquired at a source-to-detector distance of $L = 110 \: cm$, a source-to-grating distance of $L_1 = 20 \: cm$, and a grating-to-object distance of $D_{GO} = 13.5 \: cm$, for an autocorrelation length of $ACL = 57.5 \: nm$. The zoomed regions are ROI1 (yellow) and ROI2 (cyan). Note colorbar differences in original and zoomed images.}
\label{fig:anchovy_images_MPG7}
\end{figure}

\begin{figure}
\begin{subfigure}[t]{0.32\textwidth}
    \centering
    \includegraphics[keepaspectratio = true, width = \textwidth]{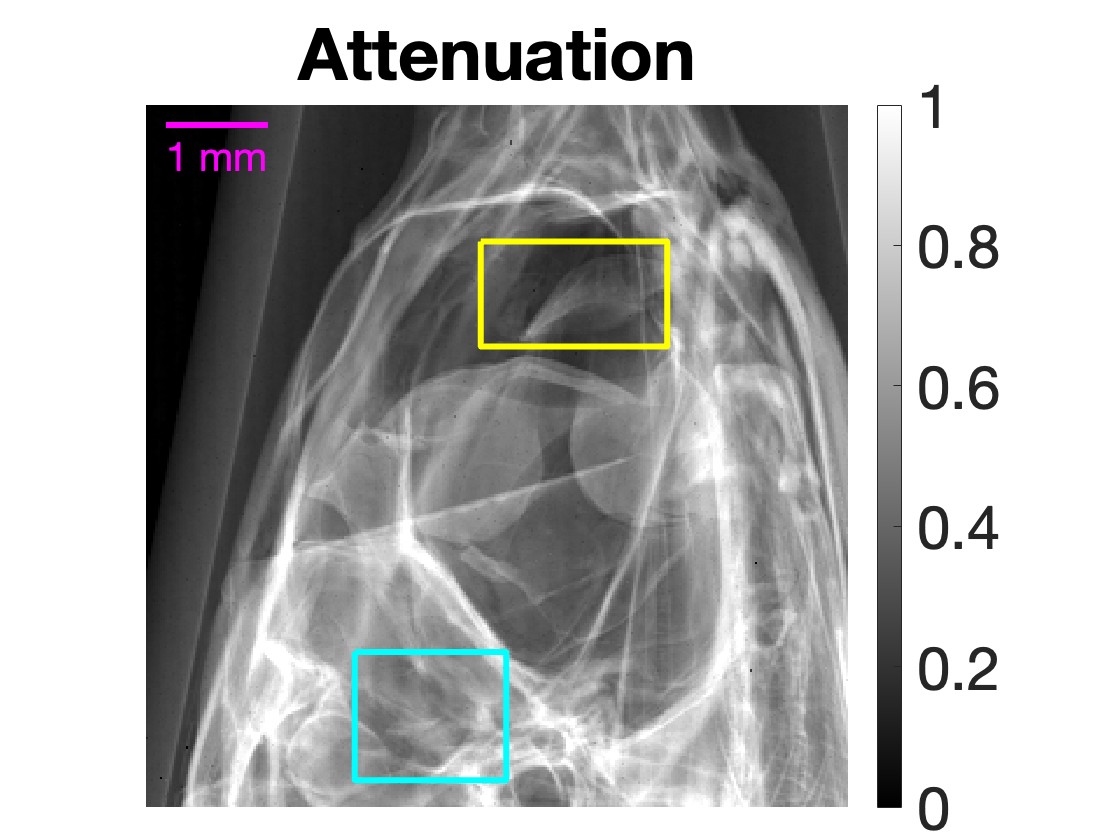}
    \caption{}
    \label{fig:anchovy_attenuation_MPG8}
\end{subfigure}
\hfill
\begin{subfigure}[t]{0.32\textwidth}
    \centering
    \includegraphics[keepaspectratio = true, width = \textwidth]{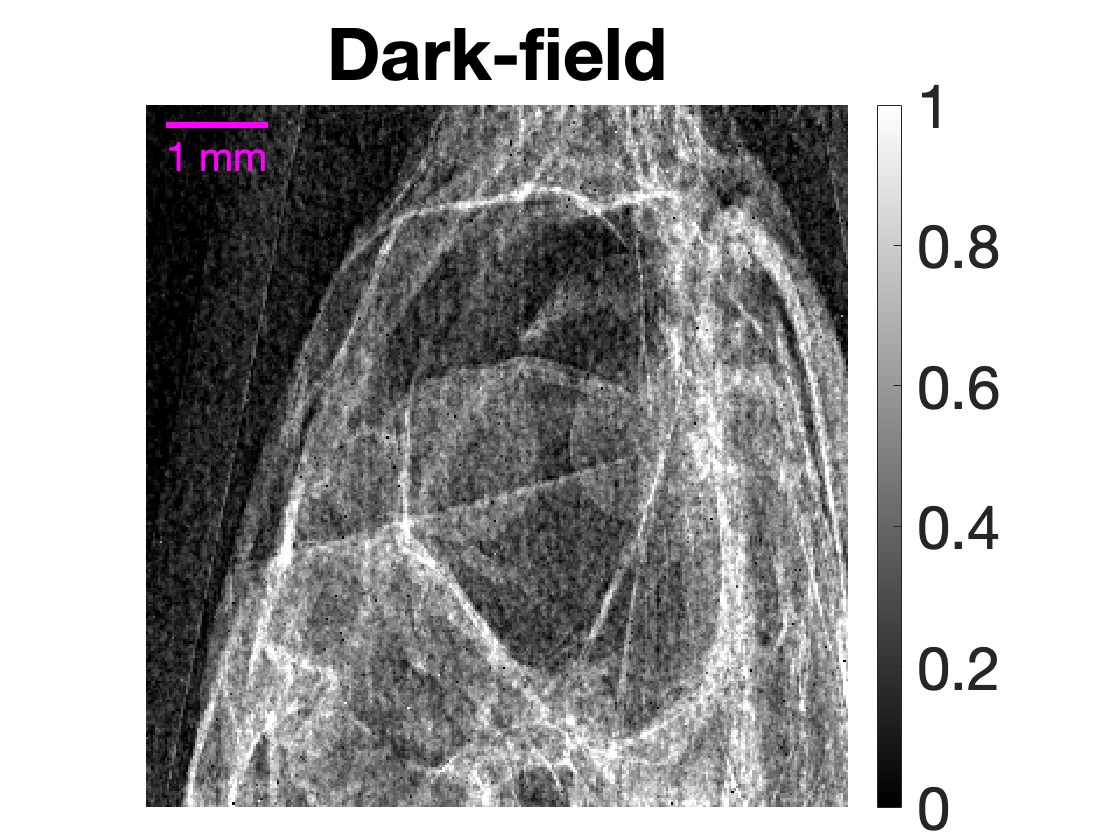}
    \caption{}
    \label{fig:anchovy_darkfield_MPG8}
\end{subfigure}
\hfill
\begin{subfigure}[t]{0.32\textwidth}
    \centering
    \includegraphics[keepaspectratio = true, width = \textwidth]{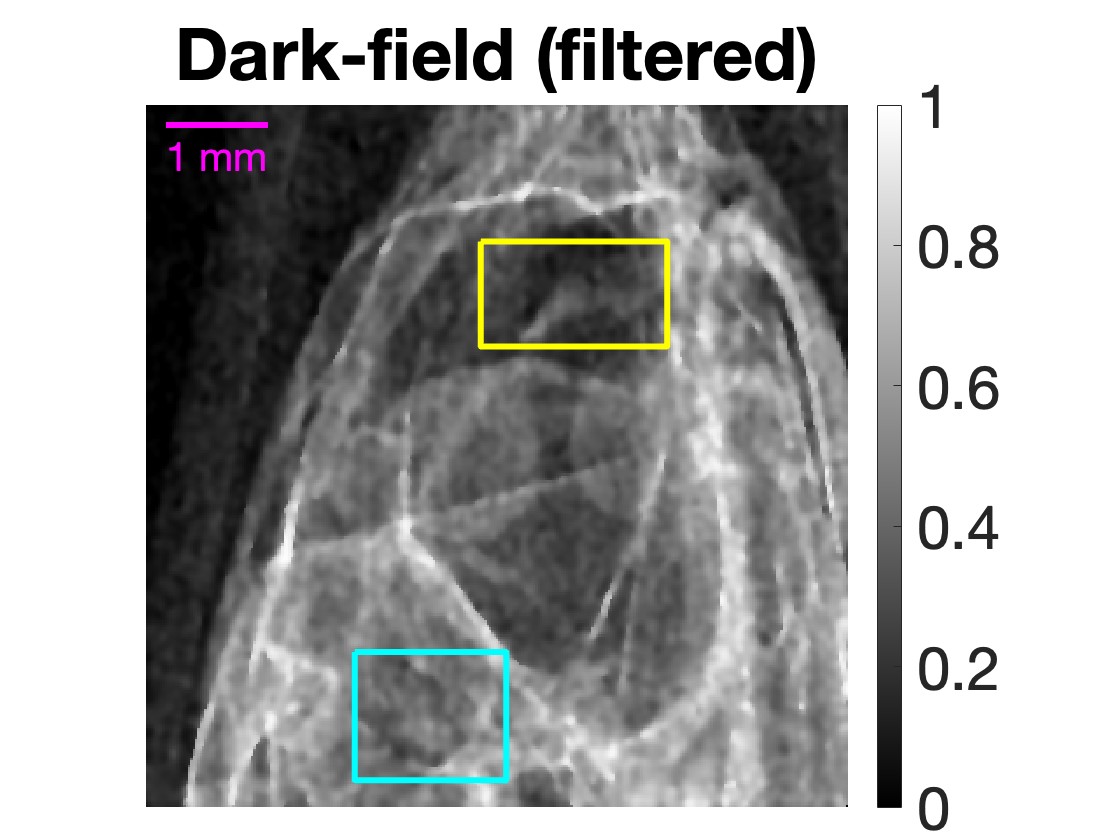}
    \caption{}
    \label{fig:anchovy_darkfield_filtered_MPG8}
\end{subfigure}

\centering
\hfill
\begin{subfigure}[t]{0.24\textwidth}
    \centering
    \includegraphics[keepaspectratio = true, width = \textwidth]{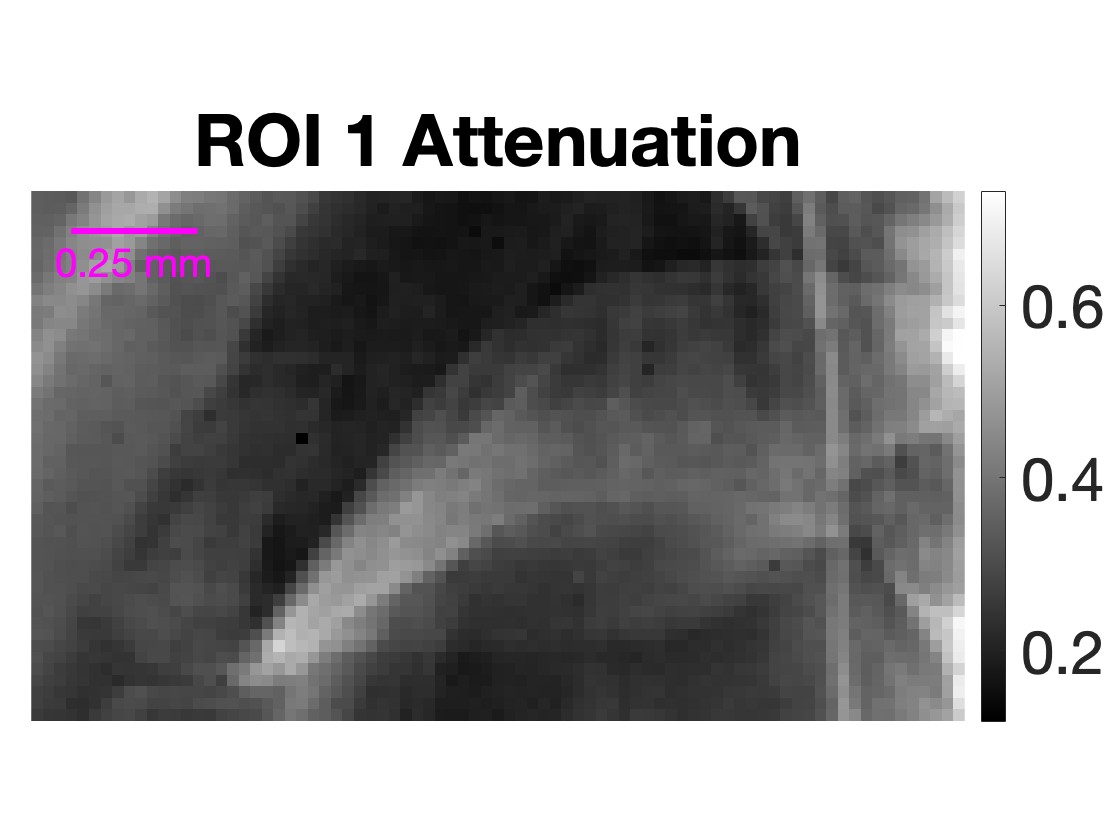}
    \caption{}
    \label{fig:anchovy_attenuation_zoomed_MPG8}
\end{subfigure}
\hfill
\begin{subfigure}[t]{0.24\textwidth}
    \centering
    \includegraphics[keepaspectratio = true, width = \textwidth]{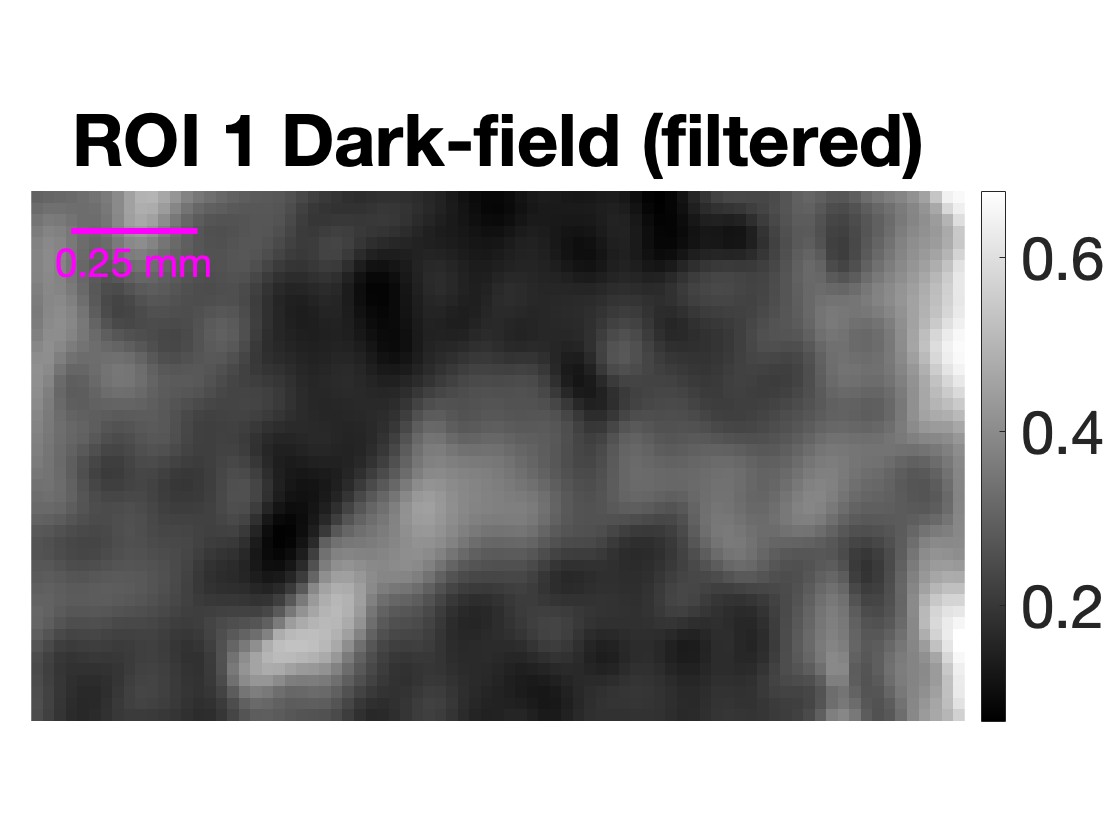}
    \caption{}
    \label{fig:anchovy_darkfield_filtered_zoomed_MPG8}
\end{subfigure}
\hfill
\begin{subfigure}[t]{0.24\textwidth}
    \centering
    \includegraphics[keepaspectratio = true, width = \textwidth]{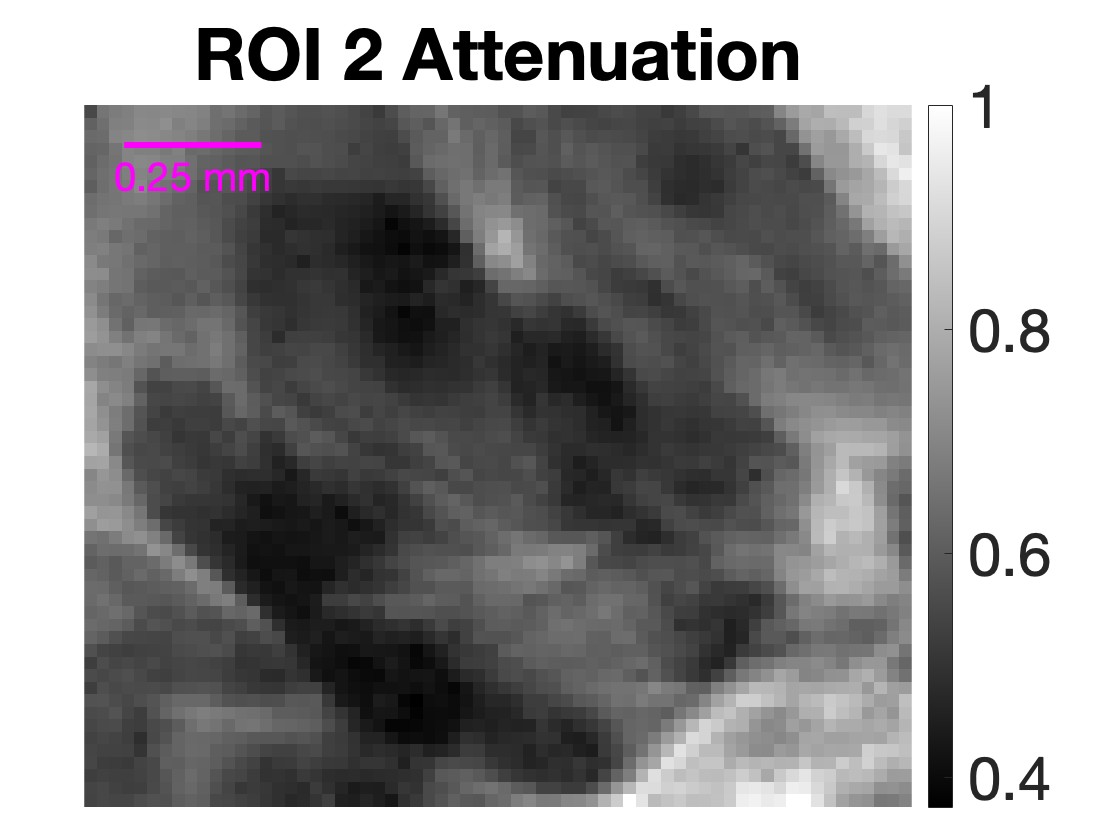}
    \caption{}
    \label{fig:anchovy_attenuation_zoomed_MPG8}
\end{subfigure}
\hfill
\begin{subfigure}[t]{0.24\textwidth}
    \centering
    \includegraphics[keepaspectratio = true, width = \textwidth]{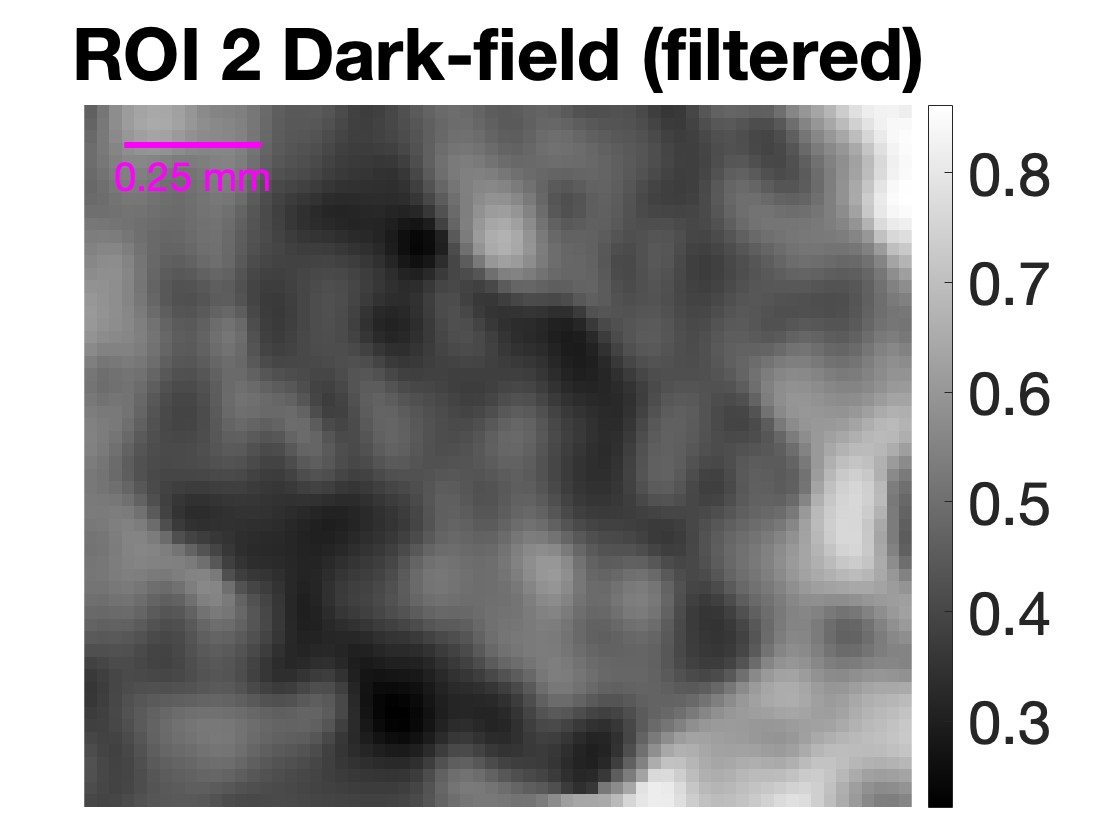}
    \caption{}
    \label{fig:anchovy_darkfield_filtered_zoomed_MPG8}
\end{subfigure}
\hspace*{\fill}

\caption{Anchovy images acquired using MPG8.  Images were acquired at a source-to-detector distance of $L = 110 \: cm$, a source-to-grating distance of $L_1 = 20 \: cm$, and a grating-to-object distance of $D_{GO} = 13.5 \: cm$, for an autocorrelation length of $ACL = 57.5 \: nm$. The zoomed regions are ROI1 (yellow) and ROI2 (cyan). Note colorbar differences in original and zoomed images}
\label{fig:anchovy_images_MPG8}
\end{figure}

Images of an anchovy were acquired with both MPG7 and MPG8, shown in Figures \ref{fig:anchovy_images_MPG7} and \ref{fig:anchovy_images_MPG8}.  At first glance, the dark-field images look like noisier versions of the attenuation images, but on closer inspection, some key differences in structure are seen, which are highlighted.  In addition to the attenuation and dark-field images, a filtered dark-field image is shown, where a 3x3 median filter was taken in addition to an anisotropic diffusion filter, with a gradient threshold of 0.2 and 4 iterations.  In the filtered image, the noise is reduced and the differences in structure are more visible.  The differences in the attenuation and dark-field images are consistent between MPG7 and MPG8.

\section{Discussion}

We presented the theory for staggered 2D gratings fabricated using a Bridge technique.  We showed that under realistic detector PSF and source blur, an approximation can be made to greatly reduce computational load known as the $l = 0$ approximation, where all y-harmonics except $l = 0$ are removed.  Using this approximation, we simulated our experimental setup for measuring the visibility of two RectMPGs, labeled MPG7 and MPG8.  In these experiments, MPG8 was shown to have a higher visibility at all distances, \textcolor{black}{which is consistent with the fact that MPG8 also has a larger difference in the height of the grating structures, $\Delta h$, as we suspect the differences seen between the two gratings result from variabilities in the height of the grating structures.  The theoretical visibility was shown to be comparable to the experimental results, with the theoretical visibility falling between the experimental visibility of the two gratings, consistent with the measured differences in the height of the grating structures, $\Delta h$, as shown in Table \ref{tab:MPG_height_measurements}. This indicates the potential for this model to be used in aiding the design of future MPGI systems.  Additional potential sources of error include improper modeling of the source spectrum or detector response function and diffraction artifacts resulting from the MPGs not exactly following the envelope function, non-rectangular grating bars, or other imperfections in the quality of the gratings. The tolerance for height control has to be determined for future applications.}

We also imaged several carbon and alumina samples and an anchovy.  The differential-phase contrast images of the ASM-385 and SAS-90 samples show the characteristic sides of the alumina spheres.  The mean normalized dark-field signal of the carbon and alumina samples was shown to follow the trend of the partial pore volume obtained from BJH adsorption and desorption differential pore volume measurements.  The dark-field anchovy images were noisy, but we did observe zones where different structures were visible in the dark-field images compared to the attenuation images, which were consistent for both MPG7 and MPG8.  For the visibility experiments, fewer phase steps were used than during the image acquisition of samples, since we were only interested in the average fringe visibility, which should not be biased by the number of phase steps if the signal-to-noise ratio is sufficiently high \cite{bib:Alegria}.

\section{Conclusion}

The modulated phase grating interferometer (MPGI) is a phase sensitive imaging system that simultaneously acquires attenuation, differential-phase contrast, and dark-field images, which show X-ray attenuation, refraction, and small angle scattering, respectively.  We have successfully modeled the 2-dimensional MPG fabricated using a Bridge technique, where staggering of the grating bars is present, and we have employed an approximation known as the $l = 0$ approximation to reduce the 2-dimensional calculation to only a single dimension, greatly reducing simulation time.  We have shown fringe visibility predictions that are comparable with experimental results and have shown that the model presented has the potential to be used to rapidly iterate when designing future MPGI systems.  We have imaged several porous carbon and alumina samples using an MPGI and have shown that the normalized dark-field signal trends well with the partial pore volume of the samples when an appropriate pore width range is used.  Lastly, we imaged a dried anchovy and showed multiple regions where unique scattering information is present.

\section{Acknowledgements}
This work is funded in part by NIH NIBIB Trail-blazer Award 1-R21-EB029026-01A1. Parts of this work were presented at the South West Regional Chapter of the American Association of Physics in Medicine (SWAAPM) in Feb 2024 and will be presented at the Graduate Research Conference, LSU in April 2024. Initial experiments were performed by author S.C. for her MSc thesis \cite{bib:SydCarrThesis}. 

\section{Conflict of Interest}
The authors have no conflicts to disclose.
\newpage

\printbibliography

\newpage

\appendix

\section{Appendix: 2D MPG Intensity Derivation}
\label{appen:2D_MPG_derivation}

\subsection{The Field Amplitude in 2 Dimensions}

For a single system geometry and energy, the angular spectrum method, \cite{bib:Goodman}, is used to derive the field amplitude, $U(x)$, for a plane wave source.  The Fresnel scaling theorem, \cite{bib:Paganin}, is used to scale the field from that of a plane wave source to that of a point-source.  Following the angular spectrum method, the angular spectrum immediately following the MPG can be found by taking the Fourier transform of Equation \ref{eq:2D_transmission},

\begin{equation}
\begin{split}
A(f_x, f_y, z = 0) &= \mathcal{F}\left(T(x,y)\right)
\\
A(f_x, f_y, z = 0) &{}={} \delta(f_x) \delta(f_y) + \alpha_x \alpha_y \sinc(\alpha_x p_x f_x) \sinc(\alpha_y p_y f_y) \left( \beta(f_x, f_y) - \gamma(f_x, f_y) \right)
\end{split}
\end{equation}

\noindent where,

\begin{dmath}
\beta(f_x, f_y) = \sum_m \sum_n \sum_l \:g_m \delta \left(f_x - \frac{m}{W} - \frac{n}{p_x}\right) \delta \left(f_y - \frac{l}{p_y}\right) \times \bigg( 1 + \exp(-j \pi \left(f_x - \frac{m}{W} \right) p_x) \exp(-j \pi f_y p_y) \bigg)
\end{dmath}

\begin{equation}
\gamma(f_x, f_y) = \sum_n \sum_l \delta \left(f_x - \frac{n}{p_x} \right) \delta \left( f_y - \frac{l}{p_y} \right) \bigg( 1 + \exp(-j \pi f_x p_x) \exp(-j \pi f_y p_y) \bigg)
\end{equation}

\noindent The angular spectrum can then be propagated to the detector,

\begin{equation}
    A(f_x, f_y, z) = A(f_x, f_y, z = 0) \: e^{jkz} \: \exp\left( -j \pi \lambda z (f_x^2 + f_y^2) \right)
\end{equation}

\noindent And the field is calculated,

\begin{equation}
\label{eq:2D_fieldAmplitude}
\begin{split}
U(x,y,z) &= \mathcal{F}^{-1}\left(A(f_x, f_y, z)\right) \\
&= e^{j k z} (1 + U_1(x,y,z) - U_2(x,y,z))
\end{split}
\end{equation}

$U_1(x,y,z)$ and $U_2(x,y,z)$ are derived from the propagation of $\beta(f_x, f_y)$ and $\gamma(f_x, f_y)$, respectively.  The $x$- and $y$-components of each term can be represented separately using matrix multiplication,

\begin{equation}
\begin{bmatrix}
    U_1(x,y,z)
    \\
    U_2(x,y,z)
\end{bmatrix}
=
\begin{bmatrix}
    U_1^{Ax}(x,z) & U_1^{Bx}(x,z)
    \\
    U_2^{Ax}(x,z) & U_2^{Bx}(x,z)
\end{bmatrix}
\times
\begin{bmatrix}
    U^{Ay}(y,z)
    \\
    U^{By}(y,z)
\end{bmatrix}
\end{equation}

\noindent where the x-components are

\begin{align}
U_1^{Ax}(x, z) {}={}& \sum_m \sum_n A_{1x}(m,n,z) \exp \left(j 2 \pi x \left( \frac{m}{W} + \frac{n}{p_x} \right) \right)
\\
U_1^{Bx}(x, z) {}={}& \sum_m \sum_n B_{1x}(m,n,z) \exp \left( j 2 \pi x \left( \frac{m}{W} + \frac{n}{p_x} \right) \right)
\\
U_2^{Ax}(x, z) {}={}& \sum_m \sum_n A_{2x}(n,z) \exp \left( j 2 \pi x \frac{n}{p_x} \right)
\\
U_2^{Bx}(x, z) {}={}& \sum_m \sum_n B_{2x}(n,z) \exp \left( j 2 \pi x \frac{n}{p_x} \right)
\\
A_{1x}(m,n,z) {}={}& \alpha_x g_m \exp \left( -j \pi \lambda z \left( \frac{m}{W} + \frac{n}{p_x} \right)^2 \right) \sinc \left( \alpha_x p_x \left( \frac{m}{W} + \frac{n}{p_x} \right) \right)
\\
A_{2x}(n,z) {}={}& \alpha_x \exp \left( -j \pi \lambda z \left( \frac{n}{p_x} \right)^2 \right)  \sinc \left( \alpha_x n \right)
\\
B_{1x}(m,n,z) {}={}& A_{1x}(m,n,z) \exp \left(-j \pi n \right)
\\
B_{2x}(n,z) {}={}& A_{2x} \exp \left(-j \pi n \right)
\end{align}

\noindent and the y-components are

\begin{align}
U_{Ay}(y, z) {}={}& \sum_l  A_y(l, y, z) \exp \left( j 2 \pi y \frac{l}{p_y} \right)
\\
U_{By}(y, z) {}={}& \sum_l B_y(l, y, z) \exp \left( j 2 \pi y \frac{l}{p_y} \right)
\\
A_y(l, z) {}={}& \alpha_y \exp \left( -j \pi \lambda z \left( \frac{l}{p_y} \right)^2 \right) \sinc \left( \alpha_y l \right)
\\
B_y(l, z) {}={}& A_y(l, y, z) \exp \left( -j \pi l \right)
\end{align}

Finally, the Fresnel scaling theorem can be applied by scaling x, y, and z by the point-source magnification factor, $M = \frac{L_1+z}{L_1}$,

\begin{gather*}
    x \longrightarrow \frac{x}{M} \\
    y \longrightarrow \frac{y}{M} \\
    z \longrightarrow \frac{z}{M}
\end{gather*}

It should be noted that there are additional phase and amplitude multiplicative factors introduced by the Fresnel scaling theorem that we do not consider for the purposes of this study, since the phase factors disappear when calculating the intensity and the amplitude factors do not affect the visibility.

\subsection{The Field Intensity and the l = 0 Approximation}

The field intensity can be calculated as simply the square of the amplitude.

\begin{equation}
    I(x,y,z) = |U(x,y,z,)|^2 = U(x,y,z)U^*(x,y,z)
\end{equation}

Since the detector will blur the y-harmonics due to the relatively small $p_y$, the detector intensity will be well approximated by the $l = 0$ harmonic, analogous to the $n = 0$ approximation present in \cite{bib:HidrovoMeyerRSI}.  To simplify this, we will represent the field amplitude from Equation \ref{eq:2D_fieldAmplitude} using only the y-harmonics (ignoring the $e^{i k z}$ which disappears in the intensity).  The $l = 0$ approximation must be taken in the \textit{intensity}, not the amplitude, but representing the field amplitude in this way will simplify the calculations.  The $l = 0$ harmonic has the primary benefit of greatly reducing the required computations, since the intensity is reduced to 1 dimension while maintaining the effects caused by the staggering of the grating bars.

\begin{equation}
    U(x,y,z) = 1 + \sum_l c(l, x, z) \exp \left( j 2 \pi y \frac{l}{M p_y} \right)
\end{equation}

\noindent where,

\begin{gather}
    c(l, x, z) = U_{Ax}(x,z) A_y(l) + U_{Bx}(x,z) B_y(l)
    \\
    U_{Ax}(x,z) = U_1^{Ax}(x,z) - U_2^{Ax}(x,z)
    \\
    U_{Bx}(x,z) = U_1^{Bx}(x,z) - U_2^{Bx}(x,z)
\end{gather}

The intensity can then be calculated:

\begin{equation}
    I(x,y,z) = 1 + \sum_l \biggl[ c(l,x,z) + c^*(-l,x,z) + d(l,x,z) \biggr] \exp \left( j 2 \pi y \frac{l}{M p_y} \right)
\end{equation}

\noindent where,

\begin{equation}
    d(l,x,z) = c(l,x,z) \star c^*(-l,x,z)
\end{equation}

The $l = 0$ approximation is then taken,

\begin{equation}
\label{eq:1D_approx_intensity}
    I(x,y,z) \approx 1 + c(0, x, z) + c^*(0, x, z) + d(0, x, z)
\end{equation}

$c(0,x,z)$ and $c^*(0,x,z)$ are easy to compute by recognizing that $A_y(0) = B_y(0) = \alpha_y$, whereas $d(0, x, z)$ can be easily computed,

\begin{dmath}
\label{eq:dl0}
d(0, x, z) = U_{Ax}U^*_{Ax} \sum_{l'} s(l') + U_{Ax}U^*_{Bx} \sum_{l'} s(l') \exp (-j \pi l') + U^*_{Ax} U_{Bx} \sum_{l'} s(l') \exp (j \pi l') + U_{Bx} U^*_{Bx} \sum_{l'} s(l')
\end{dmath}

\noindent where,

\begin{equation}
    s(l) = A_y(-l) \star A^*_y(-l) = \alpha_y^2 \sinc \left( -\alpha_y l \right)^2
\end{equation}

The $l = 0$ approximation greatly reduces the computation required and is only valid if $M*p_y$ is significantly less than the pixel size.  Notably, this does not fully remove the effect of staggering on the fringes.  The intensity profile still depends on $\alpha_y$, meaning the fringe visibility will also depend on it.

\end{document}